\documentclass[journal]{IEEEtran}
\usepackage{amsthm}
\usepackage{amsfonts}
\usepackage[top=0.43in, bottom=0.4in, left=0.45in, right=0.45in]{geometry}
\usepackage{amssymb}
\usepackage{makecell}
\ifCLASSINFOpdf
  \usepackage[pdftex]{graphicx}
  \usepackage{graphicx}
  \graphicspath{{../pdf/}{../jpeg/}{../png/}}
  \DeclareGraphicsExtensions{.pdf,.jpeg,.png}
\else
  \usepackage[dvips]{graphicx}
  \graphicspath{{../eps/}}
  \DeclareGraphicsExtensions{.eps}
\fi
\ifCLASSOPTIONcompsoc
  \usepackage[caption=false,font=normalsize,labelfont=sf,textfont=sf]{subfig}
\else
\usepackage[caption=false,font=footnotesize]{subfig}
\fi
\usepackage{multirow}
\usepackage{balance}
\usepackage{multicol}
\usepackage{epstopdf}
\usepackage{verbatim} 

\usepackage[english]{babel}
\usepackage{float}
\usepackage{xcolor}
\usepackage[linesnumbered,ruled,vlined]{algorithm2e}

\usepackage[hidelinks]{hyperref}
\usepackage{cite}
\ifCLASSINFOpdf
   \usepackage[pdftex]{graphicx}
   \else
\usepackage[dvips]{graphicx}
\fi
\usepackage[cmex10]{amsmath}
\usepackage[T1]{fontenc}
\newcommand*{\affmark}[1][*]{\textsuperscript{\dag}}

\begin{document}
\title{Subarray based Wideband Beamforming and Variational Sparse CSI Estimation for Low-Resolution MU THz MIMO Systems \vspace{-0.3 \baselineskip}}
\author{
\normalsize{Abhisha~Garg,~\IEEEmembership{Graduate Student Member,~IEEE,}  Suraj~Srivastava,~\IEEEmembership{Member,~IEEE,} Akash~Kumar and Aditya~K.~Jagannatham,~\IEEEmembership{Senior Member,~IEEE}\vspace{-2.8 \baselineskip}}

\thanks{The work is supported by IEEE SPS scholarship 2023-2025. The work of Aditya K. Jagannatham was supported in part by the Qualcomm Innovation Fellowship; in part by the Qualcomm 6G UR Gift; in part by the Arun Kumar Chair Professorship. The work of S. Srivastava was supported in part by IIT Jodhpur's Research Grant No. I/RIG/SUS/20240043; in part by Anusandhan National Research Foundation's PM-ECRG/2024/478/ENS; and in part by Telecom Technology Development Fund (TTDF) under Grant TTDF/6G/368. S. Srivastava and A. K. Jagannatham jointly acknowledge the funding support provided to ICON-project by DST and UKRI-EPSRC under India-UK Joint opportunity in Telecommunications Research and to Anusandhan National Research Foundation's Advanced Research Grant ANRF/ARG/2025/005895/ENS.

Abhisha Garg and Aditya K. Jagannatham are with the Department of Electrical Engineering, Indian Institute of Technology Kanpur, Kanpur-208016, India (e-mail: abhisha20@iitk.ac.in; adityaj@iitk.ac.in). 

Suraj Srivastava is with the Department of Electrical Engineering, Indian Institute of Technology Jodhpur, Jodhpur, Rajasthan 342030, India (e-mail: surajsri@iitj.ac.in).

Akash Kumar is with Qualcomm India Pvt. Ltd., Hyderabad, Telangana, 500081, India (email: akkum@qti.qualcomm.com).}}
\maketitle
\begin{abstract}
This work conceives a unified channel estimation and beamforming framework, formulated within the principles of variational Bayesian inference. Recognizing the limitations imposed by hardware constraints, frequency-dependent propagation effects, and the structural restrictions of partially connected architectures in the Terahertz (THz) band, we formulate a dual-wideband channel model incorporating root raised cosine (RRC) pulse shape to account its band-limited nature. To further address the nonlinear distortions introduced by low-resolution ADCs, Bussgang decomposition is employed, enabling a tractable linearized inference process. Unlike conventional techniques, the proposed method accommodates both \textit{on-grid} and \textit{off-grid angular domains}, capturing spatial sparsity with improved resolution and robustness. The multi-user (MU) Bayesian Cram{\'e}r-Rao lower bound is also derived to benchmark the performance of the proposed estimator. Moreover, the framework incorporates a true time delay (TTD)-based hybrid transceiver design that inherently compensates for the beam-squint effect; a frequency-dependent angular deviation that arises due to the fixed-phase nature of the conventional beamformer in wideband systems, thereby ensuring accurate directional alignment across all subcarriers. Extensive simulation results validate the effectiveness of the proposed variational Bayesian inference-based estimator and the TTD-enabled beamforming architecture, highlighting their robustness and performance gains under practical wideband THz system.
\end{abstract}
\IEEEpeerreviewmaketitle
\vspace{-2mm}
\section{Introduction} \label{intro}
The TeraHertz (THz) band spanning from $0.3 - 10$ THz \cite{jornet2011channel} stands out as a promising avenue to deliver ultra-high data rates and massive short-range connectivity in next-generation wireless communications. However, the new carrier frequency regime above $300$ GHz brings to the fore significant challenges in the form of severe frequency- and distance-dependent molecular absorption and path losses \cite{akyildiz2022terahertz}. In addition, the band posses frequency-dependent refractive index \cite{piesiewicz2007scattering}-\cite{piesiewicz2007properties}, which had negligible impact in the lower-frequency bands. The above degrading phenomena add to the traditional debilitating factors such as frequency-selectivity, arising due to the multipath delay spread, termed the frequency-wideband effect \cite{chuang1987effects}. Additionally, as a consequence of the large number of antenna elements required for beam steering, there arises a non-negligible propagation delay across the array aperture. This leads to differential delays in signal arrival at the various antennas, known as the \textit{spatial-wideband effect} \cite{wang2018spatial}. A general THz system that encounters both the frequency and spatial wideband effects is known as a \textit{dual-wideband system} \cite{chou2023compressed}. The \textit{beam-squint effect} \cite{wang2019beam} is observed in such a system, which arises due to the variation of the effective angle of arrival (AoA)/ angle of departure (AoD) across subcarriers, in turn affecting the array response vector. In order to avoid this loss, it is critical to incorporate the beam-squint effect in the analog beamformer/ combiner design procedure, which poses a formidable challenge. To address this impediment, this research develops novel hybrid transceiver design algorithms for MU THz MIMO systems, while considering the restricted scattering characteristics of dual-wideband THz channels \cite{wang2018spatial}. It is also worth noting that precise knowledge of the channel state information (CSI) is crucial for design of the precoder/ combiner and beam squint mitigation. Therefore, development of channel learning techniques for a dual-wideband THz channel is a worthwhile endeavor that is also addressed in this work. Moreover, conventional sparse recovery frameworks, such as orthogonal matching pursuit (OMP) \cite{karabulut2004sparse} and FOCal underdetermined system solver (FOCUSS) \cite{wipf2007empirical}, solve underdetermined systems by incorporating the sparsity constraint through equivalent penalty terms in the optimization problem. In contrast, Bayesian frameworks explicitly model the sparse structure by assigning a prior distribution and updating the information via a posterior distribution conditioned on the observations. For example, the Bernoulli-Gaussian distribution \cite{wu2016block} is widely used to represent sparse signals. Moreover, sparsity-promoting priors include \textit{Gaussian-gamma distribution} \cite{cevher2009learning}, which effectively captures the sparsity pattern of the underlying signal components. This property is particularly valuable in THz systems, where highly directional propagation leads to sparse angular support. The next section presents an overview of the prior research.
\vspace{-2mm}
\subsection{Review of existing works}\label{rev_ew}
\vspace{-1mm}
Recovering sparse signals from a single measurement vector (SMV) has been extensively explored in the literature, despite its NP-hard nature. Extending the foundational principles of SMV algorithms, multiple measurement vector (MMV) approaches have been developed to jointly process several observations of the same sparse signal \cite{cotter2005sparse}. These extensions preserve the sparsity-exploiting capabilities of SMV methods while offering improved reconstruction accuracy and robustness in more complex and practical signal recovery scenarios. Chou \textit{et al.} \cite{chou2023compressed}, proposed a multiple measurement vector least squares compressive sensing (MMV-LS-CS) based channel estimation technique for time-varying sub-THz multiple-input multiple-output orthogonal frequency division multiplexing (MIMO-OFDM) systems. The authors also introduced channel refinement and hierarchical codebook strategies to enhance path delay estimation. In his groundbreaking work, Michael E. Tipping \cite{tipping2001sparse} introduces the Relevance Vector Machine (RVM), a sparse Bayesian learning (SBL) framework for regression and classification. The proposed approach achieves high accuracy with significantly fewer basis functions than traditional methods by leveraging posterior inference to promote model sparsity and enable probabilistic prediction. While SBL provides a principled approach for sparse signal recovery by estimating the posterior distribution over weights, it often involves intractable computations in complex models. \textit{Variational Bayesian inference} addresses this by approximating the posterior with a tractable distribution, enabling efficient inference while retaining the advantages of sparsity and uncertainty quantification. Cheng \textit{et al.} \cite{cheng2017channel}, in their pioneering work, proposed a variational Bayesian inference-based channel estimation technique for MU massive MIMO systems. They capture the users' sparsity using a Gaussian mixture prior model. However, their formulation assumes a narrowband channel model, which is not suitable for wideband THz systems where frequency-dependent effects such as \textit{beam-squint} become significant. Wang \textit{et al.} \cite{wang2015novel}, proposed a wideband direction-of-arrival (DOA) framework that jointly infers angular sparsity and sub-band occupancy using a Dirichlet process prior in wideband MIMO system, and incorporates variational Bayesian algorithm to address off-grid issues.

{ Xu \textit{et al.} \cite{xu2024overcoming}, employed a variational Bayesian inference framework for mmWave MIMO channel estimation under beam squint effects. Lu \textit{et al.} \cite{lu2025adaptive}, proposed an adaptive hybrid-field channel estimation framework without prior knowledge of path components. They developed adaptive algorithms for both narrowband and wideband scenarios, incorporating preliminary near-field detection and stopping criteria to reduce computational complexity. Zhang \textit{et al.} \cite{zhang2025modular}, proposed a joint channel estimation and 3-D localization framework for modular XL-array-enabled THz systems based on the hybrid spherical-planar wave model. Further, a three-stage algorithm combining SOMP-based AoA estimation, weighted least squares (WLS)-based coarse localization, and reduced dictionary (RD)-CS-based refinement was developed to improve accuracy while reducing computational complexity. However, none of these works consider the combined impact of dual-wideband effects and low-resolution ADCs. Moreover, in the THz regime, beam squint introduces frequency-dependent variations across the array, and the interaction of the process with quantization further complicates accurate signal recovery. To address these challenges, this work proposes a novel framework for multi-user THz systems with low-resolution ADCs, where Bussgang decomposition is utilized to linearize quantized observations. Additionally, a Taylor-series-based super-resolution dictionary is developed to enable accurate sparse channel estimation, wherein higher-order expansions effectively capture off-grid effects and improve angular resolution in wideband THz channels.}

Furthermore, conventional hybrid beamforming architectures typically employ a phase shifter (PS) network to adjust the phases of transmitted signals, where the PS weights are designed at the carrier frequency and uniformly applied across the entire bandwidth \cite{lin2015adaptive}, \cite{gonzalez2018channel}, \cite{fan2015uplink} \cite{el2014spatially}. While this approach is effective in narrowband systems, where the frequency components are closely spaced, it becomes insufficient in wideband systems. This limitation arises due to the beam-squint effect, which causes the beam direction to vary with frequency. This phenomenon is especially pronounced in ultra-massive MIMO (UM-MIMO) arrays, where narrow beamwidths make the system highly sensitive to angular deviations. To mitigate this, frequency-dependent phase control is required to maintain consistent beam direction across subcarriers, which can be achieved through true time delay (TTD) elements. { The authors in \cite{dovelos2021channel} proposed a TTD-based beam squint compensation technique for sub-THz massive MIMO systems. Dai \textit{et al.} \cite{dai2022delay}, in their ground-breaking work, proposed a delay-phase precoding-based THz beamforming technique that introduces a time-delay (TD) network between the RF chains and phase shifters. They emphasize generating beams toward different physical directions and determining the corresponding time delays based on the aligned beam directions. Chang and Chiueh \cite{chang2024hybrid} extended this idea to wideband THz massive MIMO systems, introducing a TTD-based hybrid beamforming approach that employs gradient descent along with differentiable soft quantization to address quantizer discontinuities. Khan \textit{et al.} \cite{khan2026frequency}, proposed a frequency-dependent beamforming framework for ultra wideband massive MIMO systems by mapping OFDM subcarriers to distinct spatial directions, enabling full angular coverage within a single symbol. To address practical TTD limitations, a multi-stage delay-phase precoding (MSDPP) architecture was developed to mitigate hardware impairments.} However, none of these works consider low-resolution ADCs, which are critical for practical THz system design due to power constraints. To bridge this gap, the proposed framework presents a unified end-to-end THz transceiver design that integrates variational Bayesian inference-based channel estimation with TTD-enabled beamforming. Table-\ref{tab:lit_rev} boldly contrasts the salient contributions of the proposed work.
\begin{table*}
    \centering
\caption{\small Comparison of the salient contributions of the present work with existing literature} \label{tab:lit_rev}

\begin{tabular}{|l|c|c|c|c|c|c|c|c|c|c|c|c|c|c|c|c|c|}

    \hline

\textbf{Features} & \cite{wang2018spatial} &\cite{chou2023compressed} &\cite{wang2019beam}  &\cite{sha2021channel} & \cite{cheng2017channel} & \cite{xu2024overcoming} & \cite{lin2015adaptive} & \cite{gonzalez2018channel} &\cite{dovelos2021channel} &\cite{dai2022delay}&\cite{chang2024hybrid} & \cite{venugopal2017channel} &\textbf{This Paper} \\

 \hline

Sub-THz/ THz Band

&   & \checkmark &   & \checkmark  &    &   & & & \checkmark  & \checkmark &  \checkmark & & \checkmark\\

 \hline

Reflection \& Molecular absorption losses

&   & \checkmark &   &   &    &   & & & \checkmark  &  &  & & \checkmark\\

 \hline

SC-FDE

&   &  &   &  \checkmark &    &   & & &   &  &  & \checkmark & \checkmark\\
\hline

Dual-wideband channel

& \checkmark  & \checkmark &  \checkmark &   &    & \checkmark  & & & \checkmark  & \checkmark  & \checkmark & & \checkmark\\

 \hline

Off-grid dictionary

&  & &   &   &    & \checkmark  &  & \checkmark &   &  &  & & \checkmark\\

 \hline

Multi-User MIMO

&   &  &  \checkmark &   &  \checkmark  & \checkmark   &  & \checkmark &   &  &  & & \checkmark\\
\hline

AoA/AoD with GMM

&   &  &   &   &    &   & \checkmark & &   &  &  & & \checkmark\\
\hline

BCRLB

&   &  &   & \checkmark  &    &   &  & \checkmark &  \checkmark &  &  & & \checkmark\\
 \hline

\textbf{Low Resolution ADCs with beam-squint}

& &  & &  &  &  & & &   &  &  & & \checkmark\\

 \hline

\textbf{Pulse shaping filter based dual-wideband channel}

& &  &  &  &  &  & &  & &  &  & & \checkmark \\

 \hline

\textbf{Beamspace-based limited CSI feedback}

&    &   &   &   &    &   & & &  & & &  & \checkmark\\
 \hline

\textbf{Partially connected architecture in THz domain}

&    &   &   &   &    &  & &  &  & & &  & \checkmark\\
 \hline
\end{tabular}
\vspace{-1.5 \baselineskip}
\end{table*}
\vspace{-4mm}
\subsection{Contributions} \label{contri}
\begin{enumerate}
\item We commence by formulating a practical THz channel model that incorporates both frequency- and spatial-wideband effects, while also accounting for their distance-dependent characteristics. Additionally, we compare the root-raised cosine pulse shaping filter (RRC-PSF) and the rectangular pulse shaping filter (Rect-PSF) in dual-wideband channel formulations. A key challenge is the generation of distinct and realistic spatial signatures for AoA/AoD pairs using the GMM corresponding to each user, which has been overlooked in \cite{lin2015adaptive}, \cite{priebe2011aoa} and addressed in this work.
\item Exploiting angular-domain sparsity, we cast the channel estimation problem as a compressive sensing task and estimate the THz channel using \textit{variational Bayesian inference}. Notably, the Bayesian framework employed does not rely solely on the prior distribution for inference; instead, it integrates information from the likelihood function during the inference process to refine the posterior distribution. Furthermore, we incorporate both on-grid and Taylor-series-based super-resolution (TBsD) dictionaries.
\item Another key contribution of this work is the direct utilization of the estimated beamspace channel to extract the dominant AoA/AoD pairs, which are subsequently used to configure the TTD elements for beam-squint compensation. This tight integration between variational Bayesian channel learning and hybrid analog beamformer design eliminates the need for separate angle estimation modules, resulting in a low-latency, pilot-efficient solution tailored for wideband THz systems.
\item { While TTD-based architectures have been widely studied for beam-squint mitigation, existing studies largely focus on SU scenarios \cite{dovelos2021channel}, \cite{dai2022delay}, \cite{chang2024hybrid}, \cite{khan2026frequency}. However, in MU systems, the shared analog TTD network must simultaneously serve users arriving from different directions, leading to different delay requirements which results in unavoidable residual beam squint. Moreover, in wideband THz systems using low-resolution ADCs, this residual frequency-dependent array gain directly translates into non-uniform quantization distortion, since the quantization noise power scales with the ADC input variance. Therefore, this work addresses the joint delay compensation for all the users by conditioning the wideband signal prior to quantization to mitigate the joint impact of residual beam squint and coarse quantization.}
\end{enumerate}
\vspace{-4mm}
\subsection{Notation} \label{notation}
\vspace{-1mm}
The operation $\mathrm{vec}(\mathbf{ABC}) = (\mathbf{C}^T \otimes \mathbf{A})\mathrm{vec}(\mathbf{B})$ represents the vectorization of the matrix, $|\cdot|$ represents absolute operation and $\lfloor \cdot \rfloor$ represents flooring operation. The expectation operator is denoted by $\mathbb{E}\{.\}$, $\mathcal{O}(.)$ denotes the complexity order, $\Xi_N(x) = \frac{\sin(Nx/2)}{ \sin(x/2)}$ represents the unnormalized Dirichlet sinc function, while the symmetric complex Gaussian distribution is represented as $\mathcal{CN}(\boldsymbol{\mu},\boldsymbol{\mathfrak{Q}})$ where $\boldsymbol{\mu}$ represents the mean while $\boldsymbol{\mathfrak{Q}}$ represents the covariance matrix. The quantity $\mathbb{E}\{f(a,b)_{g(a),g(b)}\} = \iint f(a,b) g(a) g(b) \, da \, db$. Let $\{\mathbf{H}_u(0), \mathbf{H}_u(1), \cdots, \mathbf{H}_u(N-1)\}$ and $\{\mathbf{x}_u(0), \cdots, \mathbf{x}_u(N-1)\}$ denote a sequence of matrices and vectors, respectively. The circular convolution $\{\mathbf{r}_u(n)\}_{n=0}^{N-1}$ can be defined as
\vspace{-2mm}
\begin{align}
    \mathbf{r}_u(n) = \sum_{l=0}^{N-1}\mathbf{H}_u(l)\mathbf{x}_u[(n-l)]_N+\check{\mathbf{v}}_u(n),
\end{align}
where $[.]_N$ represents modulo-$N$ operation.
\begin{figure*}
\centering
\subfloat[]{\includegraphics[scale=0.16]{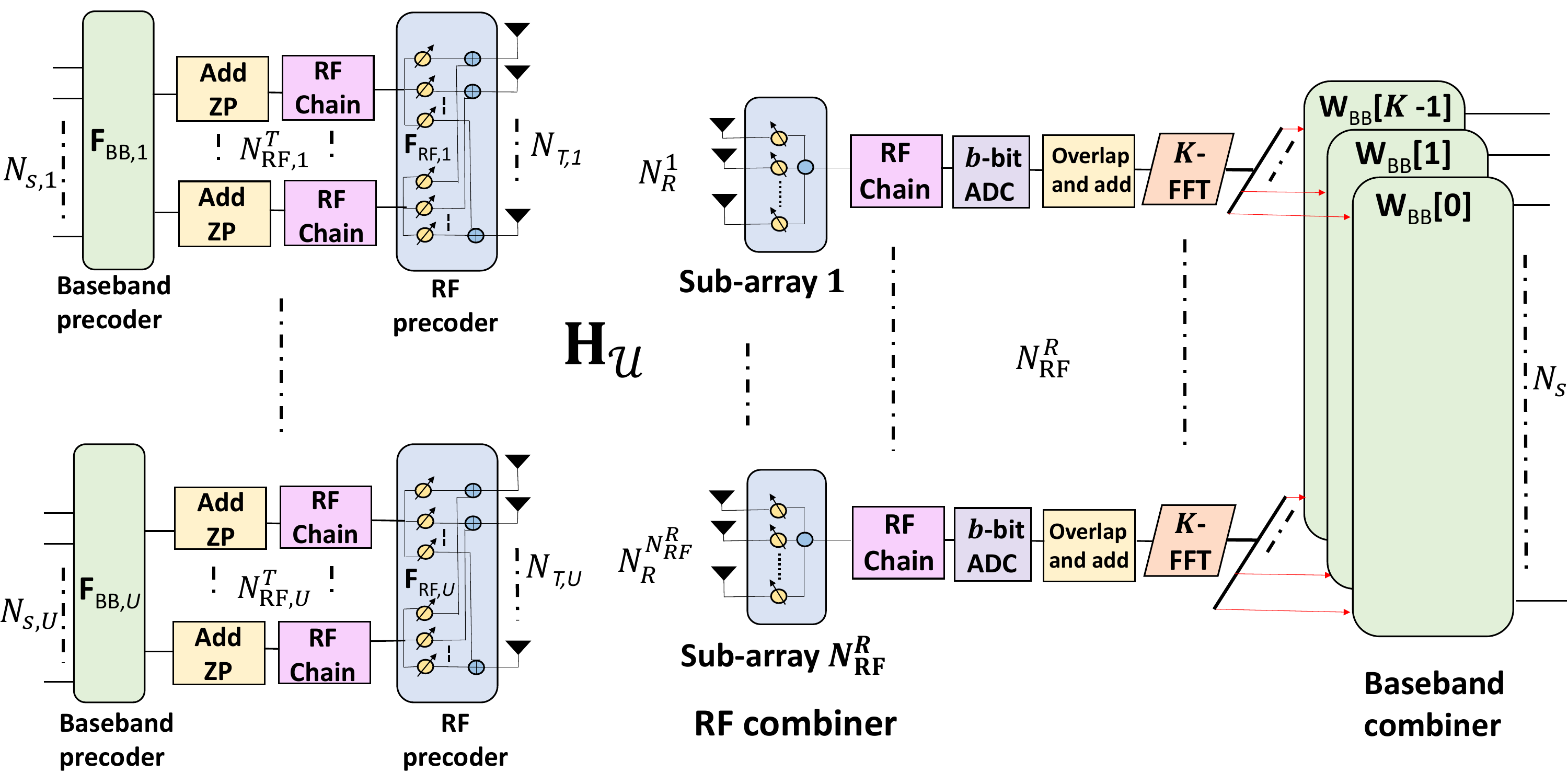}}
\hfil
\hspace{-10pt}\subfloat[]{\includegraphics[scale=0.4]{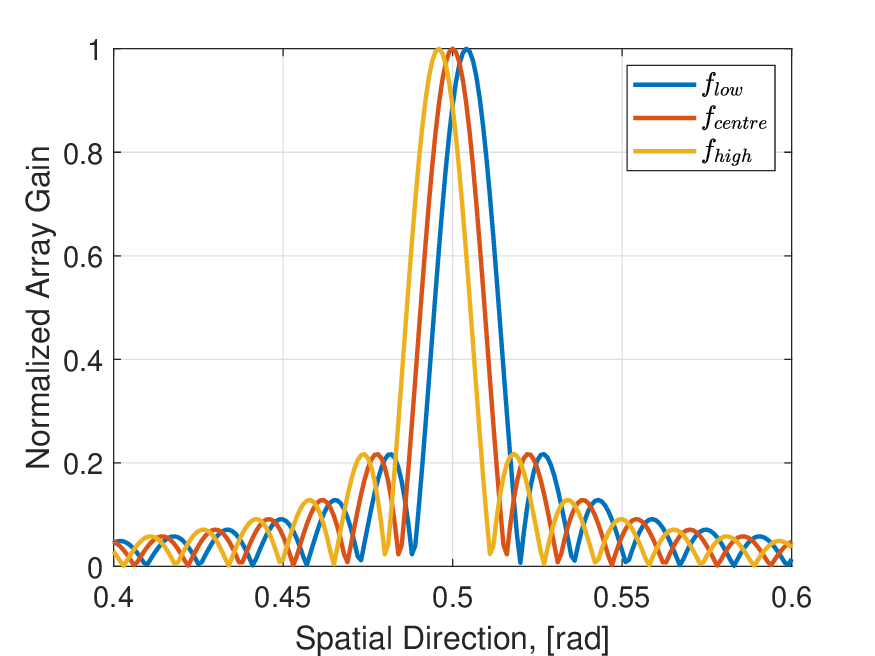}}
\vspace{-1mm}
\caption{$(a)$ Schematic diagram of SC-FDE based MU THz MIMO system $(b)$ Normalized array gain with spatial direction at $f_c = 650$ GHz, $B = 5$ GHz and $K=128$.}
\label{NAG}
\vspace{-1.5 \baselineskip}
\end{figure*}
\section{Subarray based MU-MIMO THz System Model} \label{MU_model}
Consider an uplink SC-FDMA based MU-MIMO THz system where the BS comprises $N_{\mathrm{RF}}^R$ RF chains and $N_{\mathrm{sub}}$ subarrays. Each $l$-th subarray contains $N_R^l$ antennas, serving $U$ users simultaneously. Moreover, at the transmitter, each $u$-th user has $N_{T,u}$ transmit antennas possessing $N_{\mathrm{RF},u}$ RF chain outputs and satisfies the relation $\sum_{u=1}^U N_{\mathrm{RF},u} \leq N_{\mathrm{RF}}^R = N_{\mathrm{sub}} \ll N_R$, where $N_R = \sum_{l=1}^{N_{\mathrm{RF}}^R} N_R^l$ represents the total number of antennas at the receiver, the systematic diagram for which is shown in Fig. \ref{NAG}(a). Furthermore, the total number of transmit antennas for all the users is given by $N_T = \sum_{u=1}^U N_{T,u}$, and the total number of RF chains is $N_{\mathrm{RF}}^T = \sum_{u=1}^U N_{\mathrm{RF},u}$, with each user configured in a fully-connected architecture. { Note that, the partially connected architecture is adopted in this work due to its favorable trade-offs in massive THz MIMO systems. In particular, when the number of antennas is large, the sub-connected structure achieves competitive, and in some cases superior, performance compared to the fully-connected architecture, making it well-suited for large-scale array deployments \cite[Sec.~IV-A]{du2018hybrid}.} Let $\mathbf{H}_{d,u}^l \in \mathbb{C}^{N_R^l \times N_{T,u}}, 0 \leq d < D$, represent a complex wideband THz channel corresponding to the $d$-th delay tap of $u$-th user on the $l$-th subarray, where $D$ represents the total number of taps. The frame structure consists of training and data phases. The training phase is further divided into $M$ blocks each containing $N_p$ pilot vectors. Furthermore, let $\boldsymbol{\mathfrak{P}}_{m,u}^{(p)} \in \mathbb{C}^{N_{s,u} \times 1}, 0 \leq p \leq N_p - 1$, represent the $p$-th complex pilot vector for the $m$-th block, where $N_{s,u}$ represents the number of data stream corresponding to the $u$-th user. Let the total number of data streams corresponding to all the users is given as $N_s = \sum_{u=1}^U N_{s,u}$. Prior to the transmission, the pilot vectors undergo zero-padding (ZP), where $D-1$ zeros are appended to generate a ZP block of length $K = N_p+D-1$, which is further given by $\big\{\boldsymbol{\mathfrak{P}}_{m,u}^{(p)}\Big\}_{p=0}^{K-1} = \Big\{\boldsymbol{\mathfrak{P}}_{m,u}^{(0)},\boldsymbol{\mathfrak{P}}_{m,u}^{(1)},\cdots,\boldsymbol{\mathfrak{P}}_{m,u}^{(N_p-1)},\underbrace{\mathbf{0},\cdots,\mathbf{0}}_{D-1}\Big\}$. Moreover, let $N_p - 1$ zero matrices of size $N_R^{l} \times N_{T,u}$ be appended to the channel taps $\mathbf{H}_{d,u}^{l}$, which are given by $\left\{\mathbf{H}_{p,u}^{l}\right\}_{p=0}^{K-1} = \Big\{\mathbf{H}_{0,u}^{l},\mathbf{H}_{1,u}^{l},\cdots,\mathbf{H}_{D-1,u}^{l}, \underbrace{\mathbf{0},\cdots,\mathbf{0}}_{N_p - 1} \Big\}$. Therefore, the overall channel matrix $\mathbf{H}_{p,u} \in \mathbb{C}^{N_{R} \times N_{{T},u}}$ corresponding to the $u$-th user at the BS is defined as $\mathbf{H}_{p,u} = \big[\big(\mathbf{H}_{p,u}^1\big)^T \big(\mathbf{H}_{p,u}^2\big)^T \cdots \big(\mathbf{H}_{p,u}^{N_{\mathrm{RF}}^R}\big)^T \big]^T$, where $\mathbf{H}_{p,u}^{l} \in \mathbb{C}^{N_{R}^{l} \times N_{{T},u}}$ represents the channel between the $l$-th subarray corresponding to the $u$-th user at the $p$-th time instant. Let the baseband transmit precoder at the $m$-th block and $u$-th user be given as $\mathbf{F}_{\mathrm{BB},m,u} \in \mathbb{C}^{N_{\mathrm{RF},u}^T \times N_{s,u}}$, which is cascaded with the transmit RF precoder $\mathbf{F}_{\mathrm{RF},m,u} \in \mathbb{C}^{N_{{T},u} \times N_{\mathrm{RF},u}^T}$. Therefore, the received signal vector $\Tilde{\mathbf{y}}_m(p) \in \mathbb{C}^{N_{R} \times 1}$ corresponding to the $u$-th user at the $m$-th block can be given as
\vspace{-3mm}
\begin{align}
    \mathbf{r}_m(p) = \sum_{u=1}^U \mathbf{H}_{p,u} \otimes_K \big(\mathbf{F}_{\mathrm{RF},m,u} \mathbf{F}_{\mathrm{BB},m,u} \boldsymbol{\mathfrak{P}}_{m,u}^{(p)}\big) + \Tilde{\mathbf{v}}_m(p),
\end{align}
where $\mathbf{\Tilde{v}}_m(p) \in \mathbb{C}^{N_{R} \times 1}$ represents the additive white Gaussian noise (AWGN) which follows the distribution $\mathcal{CN}(\mathbf{0}_{N_{R} \times 1}, \sigma_n^2 \mathbf{I}_{N_{{R}}})$ while $\otimes_K$ represents the circular convolution of length $K$. Note that, the ZP operation transforms the linear convolution to an equivalent circular convolution. Additionally, the received signal $\Tilde{\mathbf{y}}_m(p)$ is first passed through the analog RF combiner $\mathbf{W}_{\mathrm{RF},m} \in \mathbb{C}^{N_R \times N_{\mathrm{RF}}^R}$, followed by the quantization via low-resolution ADC, denoted as $\mathcal{Q}(.)$. { Note that, the power consumption of ADCs increases exponentially with the number of quantization bits, thereby motivating the use of low-resolution ADCs in THz systems \cite{chen2021terahertz}. Therefore, to further reduce hardware complexity and power consumption, hybrid analog-digital beamforming is jointly considered with low-resolution ADCs.} Consequently, the quantized signal after RF combining, $\mathbf{y}_m(p) \in \mathbb{C}^{N_{\mathrm{RF}}^R \times 1}$, is given by
\vspace{-3mm}
\begin{align}
    \tilde{\mathbf{y}}_m(p) = \mathcal{Q}\Big(\mathbf{W}_{\mathrm{RF},m}^H\sum_{u=1}^U\mathbf{H}_{p,u} \otimes_K & \big(\mathbf{F}_{\mathrm{RF},m,u} \mathbf{F}_{\mathrm{BB},m,u} \boldsymbol{\mathfrak{P}}_{m,u}^{(p)}\big) \notag \\ &+ \mathbf{W}_{\mathrm{RF},m}^H \Tilde{\mathbf{v}}_m(p)\Big). \label{system_model}
\end{align}
As depicted in Fig. \ref{NAG}(a), for a partially connected architecture at the BS, the RF combiner matrix can be formulated as $\mathbf{W}_{\mathrm{RF},m} = \mathrm{blkdiag}(\mathbf{w}_{\mathrm{RF},m}^1, \mathbf{w}_{\mathrm{RF},m}^2, \cdots, \mathbf{w}_{\mathrm{RF},m}^{N_{\mathrm{RF}}^R})$, where $\mathbf{w}_{\mathrm{RF},m}^l \in \mathbb{C}^{N_R^l \times 1}$ represents the analog RF vector associated with the $l$-th RF chain corresponding to the $m$-th block. Additionally, the elements of the RF precoder and combiner $|\mathbf{F}_{\mathrm{RF},u}(a,b)| = \frac{1}{\sqrt{N_{T,u}}}$ and $|\mathbf{W}_{\mathrm{RF}}(a,b)|=\frac{1}{\sqrt{N_R^l}} \: \forall a,b$ follow the unit-modulus constraint. Therefore, by applying the approximation in Eq. \eqref{system_model} and leveraging the widely adopted Bussgang decomposition for $b$-bit quantization, one can obtain
\vspace{-2mm}
\begin{align}
    \tilde{\mathbf{y}}_m(p) \approx \boldsymbol{\Lambda}\mathbf{W}_{\mathrm{RF},m}^H\sum_{u=1}^U \mathbf{H}_{p,u} & \otimes_K \big(\mathbf{F}_{\mathrm{RF},m,u} \mathbf{F}_{\mathrm{BB},m,u} \boldsymbol{\mathfrak{P}}_{m,u}^{(p)}\big) \notag \\ &+ \boldsymbol{\Lambda} \mathbf{W}_{\mathrm{RF},m}^H \Tilde{\mathbf{v}}_m(p)+\Tilde{\mathbf{v}}_q.
\end{align}
{ \underline{\textbf{Remark 1:}} Note that, to linearize the nonlinear quantization model, the Bussgang decomposition is employed, which provides good approximation for Gaussian inputs \cite{demir2020bussgang}. In the considered scenario, $\mathbf{r}_m(p)$ represents the input to the quantizer and comprises a superposition of multiple user signals undergoing linear precoding, channel propagation, and combining, in addition to additive Gaussian noise. Owing to the aggregation of multiple independent data streams and the presence of thermal noise, $\mathbf{r}_m(p)$ can be accurately approximated as a Gaussian random vector based on the \textit{central limit theorem}, thereby justifying the use of the Bussgang decomposition. Moreover, the validity of the Bussgang approximation relies on the assumption of memoryless nonlinearity, i.e., the output at any instant depends only on the current input sample and not on past inputs. However, in the presence of strong memory-based nonlinearities (e.g., nonlinear power amplifiers), this assumption underlying the Bussgang decomposition may limit its applicability \cite{goldsmith2005wireless}. This issue is further aggravated in OFDM systems due to their high peak-to-average power ratio (PAPR), which enhances nonlinear distortion and potential memory effects. In contrast, the SC-FDE framework adopted in this work, owing to its lower PAPR, mitigates such impairments and better supports the Bussgang-based linearization.}

Assuming each RF chain to be quantized independently, the quantization matrix $\boldsymbol{\Lambda} \in \mathbb{C}^{N_{\mathrm{RF}}^R \times N_{\mathrm{RF}}^R}$ can be considered to be diagonal. Moreover, the entries of $\boldsymbol{\Lambda}$ are represented as $\boldsymbol{\Lambda} = \kappa\mathbf{I}$, where $\kappa = 1-\Tilde{\xi}$ and $\Tilde{\xi}$ represents the quantization noise-to-signal power \cite{fan2015uplink}. Let $\mathbb{E}\big\{\boldsymbol{\mathfrak{P}}_{m,u}^{(p)}(\boldsymbol{\mathfrak{P}}_{m,u}^{(p)})^H\big\} = \sigma_b^2 \mathbf{I}_{N_{s,u}}$ where $\sigma_b^2$ represents the pilot power. Additionally, the quantization vector $\Tilde{\mathbf{v}}_q$ follows the distribution $\mathcal{CN}(\mathbf{0}_{N_{\mathrm{RF}}^R \times 1}, \mathbf{C}_m)$, where $\mathbf{C}_m = \kappa(1-\kappa)\mathrm{diag}(\mathbf{W}_{\mathrm{RF},m}^H\mathbf{Q}_m\mathbf{W}_{\mathrm{RF},m}+\sigma_n^2\mathbf{W}_{\mathrm{RF},m}^H\mathbf{W}_{\mathrm{RF},m})$, the derivation for which along with that for matrix $\mathbf{Q}_m$ is detailed in Appendix-\ref{appendix-a}. Once again, the diagonal error covariance matrix results from quantizing each RF chain independently which is followed from the property of Bussgang decomposition for Gaussian inputs \cite{mezghani2012capacity}. Furthermore, the values of $\tilde{\xi}$ are given in Table \ref{bit-resolution} for $b<5$ and can be approximated as $\tilde{\xi} = \frac{\pi\sqrt{3}}{2}2^{-2b}$ for $b>5$ \cite{fan2015uplink}. Let $\boldsymbol{\eta}_{m_q}(p) = \boldsymbol{\Lambda} \mathbf{W}_{\mathrm{RF},m}^H\tilde{\mathbf{v}}_m(p)+\tilde{\mathbf{v}_q} \in \mathbb{C}^{N_{\mathrm{RF}}^R \times 1}$ represent the equivalent noise. Therefore, the equivalent noise covariance matrix $\mathbf{R}_{\eta\eta} = \mathbb{E}\{\boldsymbol{\eta}_{m_q}(p) \boldsymbol{\eta}_{m_q}^H(p)\} \in \mathbb{C}^{N_{\mathrm{RF}}^R \times N_{\mathrm{RF}}^R}$ can be determined as $\mathbf{R}_{\eta\eta} = \kappa^2\sigma_n^2\mathbf{W}_{\mathrm{RF},m}^H\mathbf{W}_{\mathrm{RF},m}+\mathbf{C}_m$.
\begin{table}[t]
    \centering
    \vspace{-2mm}
    \caption{$\tilde{\xi}$ for different ADC bits $b$ \cite{fan2015uplink}}
    \label{bit-resolution}
    \begin{tabular}{|c|c|c|c|c|c|}
        \hline
        $b$ &  $1$ & $2$ & $3$ & $4$ &$5$\\ \hline
        $\tilde{\xi}$ & $0.3634$ & $0.1175$ & $0.03454$ & $0.009497$ & $0.002499$\\ \hline
    \end{tabular}
    \vspace{-2 \baselineskip}
\end{table}
After applying the $K$-point FFT on the received signal as $\left\{{\mathbf{y}}_m[k]\right\}_{k=0}^{K-1} = \mathrm{FFT}\big(\big\{{\mathbf{y}}_m(p)\big\}_{p=0}^{K-1}\big)$, one can obtain its frequency domain equivalent $\mathbf{y}_m[k] \in \mathbb{C}^{N_{\mathrm{RF}}^R \times 1}$ as
\vspace{-5mm}

\small
\begin{align}
    \tilde{\mathbf{y}}_m[k] \approx \boldsymbol{\Lambda}\mathbf{W}_{\mathrm{RF},m}^H\sum_{u=1}^U \mathbf{H}_u[k]\mathbf{F}_{\mathrm{RF},m,u}\mathbf{F}_{\mathrm{BB},m,u}\boldsymbol{\mathfrak{P}}_{m,u}[k] + \boldsymbol{\eta}_{m_q}[k], \notag
\end{align}
\normalsize
where the quantity $\boldsymbol{\mathfrak{P}}_{m,u}[k] \in \mathbb{C}^{N_{s,u} \times 1}$ represents the equivalent transmit pilot vector corresponding to the $m$-th block at the $k$-th subcarrier for the $u$-th user, which is given by the $ \mathrm{FFT}\big(\big\{\boldsymbol{\mathfrak{P}}_{m,u}^{(p)}\big\}_{p=0}^{K-1}\big)$. Similarly, the quantity $\boldsymbol{\eta}_{m_q}[k] \in \mathbb{C}^{N_{\mathrm{RF}}^R \times 1}$ represents the FFT of $\boldsymbol{\eta}_{m_q}(p)$. The received signal $\mathbf{r}_m[k] \in \mathbb{C}^{N_s \times 1}$ after baseband combining with $\mathbf{W}_{\mathrm{BB},m}[k] \in \mathbb{C}^{N_{\mathrm{RF}}^R \times N_s}$ is given as
\begin{align}
    \mathbf{y}_m[k] \approx \mathbf{W}_{\mathrm{BB},m}^H[k] \boldsymbol{\Lambda} \mathbf{W}_{\mathrm{RF},m}^H & \mathbf{H}_\mathcal{U}[k]\Tilde{\mathbf{F}}_{\mathrm{RF},m}\Tilde{\mathbf{F}}_{\mathrm{BB},m}\Tilde{\boldsymbol{\mathfrak{P}}}_{m}[k] + \notag \\ & \mathbf{W}_{\mathrm{BB},m}^H[k] \boldsymbol{\eta}_{m_q}[k], \label{received}
\end{align}
where $\mathbf{H}_\mathcal{U}[k] = \left[\mathbf{H}_1[k] \: \mathbf{H}_2[k] \cdots \mathbf{H}_U[k] \right] \in \mathbb{C}^{N_R \times N_T}$ represents the concatenated channel for all the users. { Therefore, for a partially connected wideband THz MU hybrid MIMO system, Eq. \eqref{received} characterizes the considered end-to-end system model while accounting for the effects of low-resolution ADCs.} Moreover, the overall transmit precoder $\Tilde{\mathbf{F}}_{\mathrm{RF},m} \in \mathbb{C}^{N_{T} \times N_{\mathrm{RF}}^T}$ and baseband precoder $\Tilde{\mathbf{F}}_{\mathrm{BB},m} \in \mathbb{C}^{N_{\mathrm{RF}}^T \times N_s}$ for all the users in the $m$-th block can be expressed as $\Tilde{\mathbf{F}}_{\mathrm{RF},m} = \mathrm{blkdiag}\left(\mathbf{F}_{\mathrm{RF},m,1} \: \mathbf{F}_{\mathrm{RF},m,2} \cdots \mathbf{F}_{\mathrm{RF},m,U} \right)$ and $\Tilde{\mathbf{F}}_{\mathrm{BB},m} = \mathrm{blkdiag}\left(\mathbf{F}_{\mathrm{BB},m,1} \: \mathbf{F}_{\mathrm{BB},m,2} \cdots \mathbf{F}_{\mathrm{BB},m,U} \right)$, respectively. Furthermore, the stacked pilot vector $\Tilde{\boldsymbol{\mathfrak{P}}}_m[k] \in \mathbb{C}^{N_s \times 1}$, representing all users at the $m$-th block, is given as $\Tilde{\boldsymbol{\mathfrak{P}}}_m[k] = \big[\boldsymbol{\mathfrak{P}}_{m,1}^T[k] \: \boldsymbol{\mathfrak{P}}_{m,2}^T[k] \cdots \boldsymbol{\mathfrak{P}}_{m,U}^T[k]\big]$.

After applying the $\mathrm{vec}(.)$ operator as described in Section-\ref{notation}, Eq. \eqref{received} can be written as
\vspace{-2mm}
\begin{align}
    \mathbf{y}_m[k] \approx &\underbrace{\big(\Tilde{\boldsymbol{\mathfrak{P}}}_m^T[k]\Tilde{\mathbf{F}}_{\mathrm{BB},m}^T\Tilde{\mathbf{F}}^T_{\mathrm{RF},m}\big) \otimes \left(\mathbf{W}_{\mathrm{BB},m}^H[k] \boldsymbol{\Lambda}\mathbf{W}_{\mathrm{RF},m}^H\right)}_{\tilde{\mathbf{\Psi}}_{\mathcal{U},m}[k] \in \mathbb{C}^{N_s \times N_{T} N_{R}}} \times \notag \\ & \quad\quad\quad\quad\quad\quad\quad\quad\quad \underbrace{\mathrm{vec}(\mathbf{H}_\mathcal{U}[k])}_{\mathbf{h}_\mathcal{U}[k]} + \boldsymbol{\eta}_m[k], \label{sys-mod}
\end{align}
where $\mathbf{h}_\mathcal{U}[k] \in \mathbb{C}^{N_{R} N_{T} \times 1}$ represents the vectorized concatenated channel for all the users while $\boldsymbol{\eta}_m[k] = \mathbf{W}_{\mathrm{BB},m}^H[k]\boldsymbol{\eta}_{m_q}[k] \in \mathbb{C}^{N_s \times 1}$ represents the combined effective noise, which obeys the distribution $\mathcal{CN}(\mathbf{0}_{N_s \times 1},\mathbf{D}_m[k])$, where $\mathbf{D}_m[k] = \mathbf{W}_{\mathrm{BB},m}^H[k]\mathbf{R}_{\eta\eta} \mathbf{W}_{\mathrm{BB},m}[k]$. Moreover, the quantity $\tilde{\mathbf{\Psi}}_{\mathcal{U},m}[k]$ represents the combined sensing matrix for all the users at the $m$-th block and $k$-th subcarrier. Additionally, the sum spectral efficiency for the $m$-th block can be given as
\vspace{-3mm}
\begin{align}
    \mathrm{SE} = \frac{1}{K} \sum_{k=0}^{K-1} \mathrm{log}_2 \big|\mathbf{I}_{N_s} + \frac{1}{N_s} \mathbf{D}_m^{-1}[k] \mathbf{G}_m[k]\big|,
\end{align}
where $\mathbf{G}_m[k] = \mathbf{W}_m^H[k]\mathbf{H}_\mathcal{U}[k] \mathbf{F}_m \mathbf{F}_m^H\mathbf{H}_\mathcal{U}^H[k]\mathbf{W}_m[k]$, $\mathbf{W}_m[k] = \mathbf{W}_{\mathrm{RF},m}\boldsymbol{\Lambda}\mathbf{W}_{\mathrm{BB},m}[k]$ while $\mathbf{F}_m = \Tilde{\mathbf{F}}_{\mathrm{RF},m}\Tilde{\mathbf{F}}_{\mathrm{BB},m}$. The next section will detail the formation of partially connected dual-wideband channel and its sparse representation. 
\vspace{-2mm}
\section{MU-THz channel model and sparse formation} \label{Channel_mod}
\vspace{-1mm}
The generalized array steering vector $\mathbf{a}^{l}_{R}(\theta,f) \in \mathbb{C}^{N_R^l \times 1}$ for a partially connected architecture at the $l$-th subarray can be given as
\vspace{-2mm}
\begin{align}
    \mathbf{a}_{R}^{l}\left(\theta,f\right) = \frac{1}{\sqrt{N_R^l}} \big[1, e^{-j \frac{2 \pi}{\lambda} d_c \sin \theta}, \cdots, e^{-j \frac{2 \pi}{\lambda} d_c (N_{R}^{l}-1) \sin \theta} \big]^T, \label{arr_res_fc}
\end{align}
where the quantity $\lambda = \frac{c}{f}$ represents the wavelength, $d_c = \frac{c}{2f_c}$ denotes the inter-antenna-spacing and $f_c$ is the central carrier frequency. Note that, the approximation $d_c \approx \frac{\lambda}{2}$ holds only in the narrowband case, i.e., when $B \ll f_c$, where $B$ denotes the bandwidth. Therefore, the effective spatial AoA at the $k$-th subcarrier, can be defined as
\vspace{-2mm}
\begin{align}
    \theta_k = \mathrm{arcsin}(\vartheta_k \sin\theta),
\end{align}
where $\vartheta_k = \frac{f_k}{f_c}$ represents the relative frequency of the subcarrier and $f_k \triangleq f_c + \big(k-\frac{K+1}{2}\big)\frac{B}{K}$ denotes the $k$-th subcarrier frequency. Moreover, the modified array steering vector after incorporating the spatial wideband effect can be given as
\vspace{-4mm}

\small
\begin{align}
    \Tilde{\mathbf{a}}_{R}^{l}\left(\theta,f_k\right) = \frac{1}{\sqrt{N_{R}^{l}}}\big[1, e^{-j \pi \frac{f_k}{f_c} \sin \theta}, \cdots, e^{-j \pi \left(N_{R}^{l}-1\right) \frac{f_k}{f_c} \sin \theta} \big]. \label{arrayresponsevector}
\end{align}
\normalsize
Therefore, the normalized array gain \cite{dai2022delay} is given as
\vspace{-2mm}
\begin{align}
\tilde{\Omega} = |(\tilde{\mathbf{a}}_R^l(\theta_k,f_k))^H\tilde{\mathbf{a}}_R^l(\theta,f_c)|^2, \label{normalizedgain}
\end{align}
which is plotted in Fig. \ref{NAG}(b). For a wideband THz system i.e., $\Tilde{\mathbf{a}}_{R}^{l}\left({\theta}_k,f_k \right) \neq \Tilde{\mathbf{a}}_{R}^{l}(\theta,f_c)$ leads to $\tilde{\Omega} \neq 1$; and therefore the beams generated by $\Tilde{\mathbf{a}}_{R}^{l}(\theta,f_k)$ point in different physical directions $\theta_k$, leading to the \textit{beam-squint effect}, as illustrated in Fig. \ref{NAG}. This effect causes different subcarriers to observe distinct AoA/AoDs for the same physical path. Moreover, the high-gain beams exhibit significant deviation from the intended direction across subcarriers. The maximum array gain is achieved only when $\theta_k \vartheta_k - \theta = 0$, the derivation for which is detailed in Appendix-\ref{appendix-b}.

The THz MIMO channel can be aggregated as the sum of line of sight (LoS) and non-LoS (NLoS) components given as $\mathbf{H}_u^{l}[k] = \mathbf{H}_u^{l,\mathrm{LoS}}[k] + \mathbf{H}_u^{l,\mathrm{NLoS}}[k]$ where 
\begin{align}
    \mathbf{H}_u^{l,\mathrm{LoS}}[k] = \sqrt{N_{{T},u} N_{R}^{l}} \alpha(f_k,d)\beta_\tau \mathfrak{D}_{T,u}\mathfrak{D}_R^l\tilde{\mathbf{a}}_{R}^l(\theta,f_k)\tilde{\mathbf{a}}_u^H(\varphi,f_k),
\end{align}\label{LoS_channel}
\vspace{-7mm}
\begin{align}
    \mathbf{H}_u^{l,\mathrm{NLoS}}[k] = & \sqrt{\frac{N_{T,u} N_R^l}{N_{\mathrm{NLoS}}N_{\mathrm{ray}}}} \sum_{\ell=1}^{N_{\mathrm{NLoS}}} \sum_{\jmath=1}^{N_{\mathrm{ray}}}\alpha_{\ell,\jmath}(f_k,d_{\ell,\jmath}) \notag \\ & \beta_{\tau_{\ell,\jmath}}\mathfrak{D}_{T,u}\mathfrak{D}_R^l\tilde{\mathbf{a}}_R^l(\theta_{\ell,\jmath},f_k)\tilde{\mathbf{a}}_u^H(\varphi_{\ell,\jmath},f_k),
\end{align}\label{NLoS_channel}
where $\beta_{\tau_{\ell,\jmath}} = \sum_{\imath=0}^{K-1}p(\imath T_s - \tau_{\ell,\jmath})e^{-j\frac{2 \pi k_{\imath}}{K}}, \forall \: k,\imath$. The quantities $p(.), T_s, \tau_{(.)}, \theta_{(.)}, \varphi_{(.)}, N_{\mathrm{NLoS}}, N_{\mathrm{ray}}$ denote the RRC-PSF, sampling time, delay, AoA, AoD, number of NLoS components and number of diffused rays, respectively, where $(.)$ represents the LoS/ NLoS complex path which are detailed in \cite{garg2024angularly}. The next section will discuss the on-grid and off-grid dictionary based sparse channel formulation.
\vspace{-4mm}
\subsection{On-grid array manifold and dictionary formulation}
\vspace{-1mm}
Let $G_{T,u}$ and $G_R^l$ represent the number of transmit and receive angular bins. Further, let $\Theta_R^l$ and $\Phi_{T,u}$ represent the receive and transmit directional sines which are given as
\begin{align}
    \Phi_{T,u} = \{\varphi_{t,u}: \sin(\varphi_{t,u}) = \frac{2}{G_{T,u}}(t-1)-1, 1 \leq t \leq G_{T,u}\}, \notag
\end{align}
\vspace{-6mm}
\begin{align}
    \Theta_R = \{\theta_R: \sin(\theta_R) = \frac{2}{G_R^l}(r-1)-1, 1\leq r \leq G_R^l \}.
\end{align}
Therefore, the extended virtual channel representation corresponding to the $l$-th subarray can be given as $\mathbf{H}_u^l[k] = \mathbf{A}_R^l(\Theta_R,f_k)\mathbf{H}_{b,u}^l[k]\mathbf{A}_{T,u}^H(\Phi_{T,u},f_k)$, where $\mathbf{H}_{b,u}^l[k] \in \mathbb{C}^{N_R^l \times N_{T,u}}$ represents the beamspace channel frequency response matrix (CFR) corresponding to $\mathbf{H}_u^l[k]$. The quantities $\mathbf{A}_R^l(\Theta_R,f_k) \in \mathbb{C}^{N_R^l \times G_R^l}$ and $\mathbf{A}_{T,u}(\Phi_{T,u},f_k) \in \mathbb{C}^{G_{T,u} \times N_{T,u}}$ represent the receive and transmit array manifold dictionaries which are further given
\begin{align}
    \mathbf{A}_R^l(\Theta_R,f_k) = [\tilde{\mathbf{a}}_R^l(\theta_1,f_k) \: \tilde{\mathbf{a}}_R^l(\theta_2,f_k) \cdots \tilde{\mathbf{a}}_R^l(\theta_{G_R^l},f_k)], \notag
\end{align}
\vspace{-4mm}
\begin{align}
    \mathbf{A}_{T,u}(\Phi_{T,u},f_k) = [\tilde{\mathbf{a}}_{T,u}(\varphi_{1,u},f_k)  \cdots \tilde{\mathbf{a}}_{T,u}(\varphi_{G_{T,u}},f_k)].
\end{align}
Since, the AoAs/AoDs are inherently continuous and may not align precisely with the discretized grid points, this lead to \textit{basis mismatch}, which undermines the conventional sparsity assumption. Although one can construct a finer quantized dictionary to mitigate basis mismatch, this approach incurs significantly higher computational complexity. Therefore, the next section presents a TBsD formulation.
\vspace{-4mm}
\subsection{Taylor series based super-resolution dictionary (TBsD) formulation}
The antenna array $\tilde{\mathbf{a}}_R^l(\theta,f_k)$ is a continuous and smooth function, and therefore, it can be approximated by linearly combining $\tilde{\mathbf{a}}_R^l(\theta_r,f_k)$ and its derivatives via a first-order Taylor expansion given as
\vspace{-5mm}

\small
\begin{align}
    \tilde{\mathbf{a}}_R^l(\theta,f_k) = \tilde{\mathbf{a}}_R^l(\theta_r,f_k) + (\theta-\theta_r) \frac{\partial \tilde{\mathbf{a}}_R^l(\theta,f_k)}{\partial \theta}\big|_{\theta_r} + \mathcal{O}((\theta-\theta_r)^2),
\end{align} \label{tay_app}
\normalsize
where $\theta_r = 2 \pi \frac{(r-1)}{G_R^l}$ represents the grid-point with minimal distance to $\theta$. Therefore, the modified dictionary matrix will contain the grid points from the quantized array along with their derivatives given as
\begin{align}
    \boldsymbol{\digamma}_R^l(\theta_r,f_k) = &[\tilde{\mathbf{a}}(\theta_1,f_k),\cdots ,\tilde{\mathbf{a}}(\theta_{G_R^l},f_k),{\mathbf{b}}(\theta_1,f_k),  \cdots,\mathbf{b}(\theta_{G_R^l,f_k})], \notag
\end{align}
where $\mathbf{b}(\theta_r,f_k) = \frac{\partial \tilde{\mathbf{a}}_R^l(\theta, f_k)}{\partial \theta}\big|_{\theta_r} \in \mathbb{C}^{2G_R^l \times 1}$. Furthermore, an interpolator $\boldsymbol{\mathfrak{t}}_{R}^l \in \mathbb{C}^{2G_R^l \times 1}$ can be defined, which contains the offset angles $\triangle \theta$ expressed as
\vspace{-2mm}
\begin{align}
    \boldsymbol{\mathfrak{t}}_R^l = \big[\underbrace{1,\cdots,1}_{G_R^l},\underbrace{\triangle \theta, \cdots, \triangle \theta}_{G_R^l}\big]^T,
\end{align}
where $|\triangle \theta| \leq \frac{\pi}{G_R^l}$. Therefore, the TBsD at the BS can be given as
\vspace{-2mm}
\begin{align}
    & \tilde{\mathbf{A}}_R^l(\Theta_R,f_k) = \boldsymbol{\digamma}_R^l\boldsymbol{\mathfrak{t}}_R^l = \notag \\ & [\tilde{\mathbf{a}}(\theta_1,f_k),\cdots,\tilde{\mathbf{a}}(\theta_{G_R^l},f_k),\mathbf{b}(\theta_1,f_k),\cdots,\mathbf{b}(\theta_{G_R^l},f_k)]^T.
\end{align}
Note that the TBsD is invariant to the order of operations, thus performing subarray-based stacking followed by linearization yields identical results to first linearizing and subsequently stacking across subarrays. For simplicity and clarity of presentation, we first linearize using the TBsD and then stack across all subarrays.\\
{ \underline{Validity range:} To ensure accurate angular representation, the angular offset $\triangle \theta$ must satisfy two requirements. First, to guarantee full angular coverage, the grid resolution must satisfy $\triangle \theta \approx \frac{\pi}{G_R^l}$ \cite{bishnu2017sparse},\cite{zhang2024sparse}. Second, for the first-order Taylor approximation to remain valid, the angular deviation must be sufficiently small such that the induced phase variation within each subarray is negligible. This imposes the condition $|\triangle \theta|\ll \frac{\lambda_k}{2 \pi d_c (N_R^l -1)}$ \cite{kay1993statistical}. Combining these constraints yields the following requirement on the grid size $G_R^l \geq \frac{2 \pi^2 d(N_R^l -1)}{\lambda_k}$. Under this condition, the grid resolution becomes sufficiently fine such that the selected grid point $\theta_r$ lies in close proximity to the true angle $\theta$, thereby ensuring that the angular deviation $|\theta-\theta_r|$ remains sufficiently small. Consequently, the higher-order terms in the Taylor expansion can be safely neglected, and the first-order approximation in Eq. \eqref{tay_app} is adopted.} Similarly, the TBsD at the transmitter for $u$-th user is
\vspace{-2mm}
\begin{align}
    \tilde{\mathbf{A}}_{T,u}(\Phi_{T,u},f_k) & = [\tilde{\mathbf{a}}_{T,u}(\phi_1,f_k), \cdots, \tilde{\mathbf{a}}_{T,u}(\phi_{G_{T,u}},f_k), \notag \\ & \mathbf{b}_{T,u}(\phi_1,f_k),\cdots,\mathbf{b}_{T,u}(\phi_{G_{T,u}},f_k)].
\end{align}
Therefore, the extended virtual channel model\footnote{Note that, for clarity of exposition, the derivations have been presented for the on-grid case. The same procedure can be extended to the off-grid scenario by replacing $\mathbf{A}_R^l(\Theta_R,f_k) \rightarrow \tilde{\mathbf{A}}_R^l(\Theta_R,f_k)$ and $\mathbf{A}_{T,u}(
\mathbf{\Phi}_{T,u}) \rightarrow \tilde{\mathbf{A}}_{T,u}(
\mathbf{\Phi}_{T,u})$, respectively.} corresponding to each user for all the subarrays can be given as $\mathbf{H}_u[k] = \mathbf{A}_R(\Theta_R,f_k) \mathbf{H}_{b,u}[k] \mathbf{A}_{T,u}^H(\Phi_{T,u},f_k)$, where $\mathbf{A}_R(\Theta_R,f_k) \in \mathbb{C}^{N_R \times G_R}$ is expressed as
\vspace{-2mm}
\begin{align}
     \mathbf{A}_R(\Theta_R,f_k) = \mathrm{blkdiag}(\mathbf{A}_R^1(\Theta_R,f_k) \cdots \mathbf{A}_R^{N_{\mathrm{RF}}^R}(\Theta_R,f_k)),
\end{align}
and $G_R = \sum_{l=1}^{N_{\mathrm{RF}}^R} G_R^l$. Let $\boldsymbol{\Sigma}_u [k] \in \mathbb{C}^{N_R N_{T,u} \times G_R G_{T,u}}$ represent the \textit{sparsifying-dictionary} matrix defined as $\boldsymbol{\Sigma}_u [k] = \mathbf{A}_{T,u}^*(\Phi_{T,u},f_k) \otimes \mathbf{A}_R(\Theta_R,f_k)$. Therefore, the vectorized CFR corresponding to all the users is given as $\mathbf{h}_{\mathcal{U}}[k] = \mathrm{blkdiag}\underbrace{(\boldsymbol{\Sigma}_1[k] \: \boldsymbol{\Sigma}_2[k] \cdots \boldsymbol{\Sigma}_U[k])}_{\boldsymbol{\Sigma}_{\mathcal{U}}[k]} \underbrace{[\mathbf{h}_{b,1}^T[k] \: \mathbf{h}_{b,2}^T[k] \: \mathbf{h}_{b,U}^T[k]]^T}_{\mathbf{h}_{b,\mathcal{U}}[k]}$, where $\boldsymbol{\Sigma}_{\mathcal{U}}[k] \in \mathbb{C}^{N_RN_T \times G_RG_T}$ represents the joint sparsifying dictionary and $\mathbf{h}_{b,\mathcal{U}}[k] \in \mathbb{C}^{G_R G_T \times 1}$ denotes the joint beamspace output. Therefore, the joint sparse signal model $\mathbf{y}_{m}[k] \in \mathbb{C}^{N_s \times 1}$ for all the users at $m$-th block can be given
\begin{align}
    \mathbf{y}_m[k] \approx \underbrace{\tilde{\mathbf{\Psi}}_{\mathcal{U},m}[k]\boldsymbol{\Sigma}_{\mathcal{U}}[k]}_{\mathbf{\Psi}_{\mathcal{U},m}[k]}\mathbf{h}_{b,\mathcal{U}}[k] + \boldsymbol{\eta}_m[k],
\end{align}
where $\mathbf{\Psi}_{\mathcal{U},m}[k] \in \mathbb{C}^{N_s \times G_RG_T}$ represents the equivalent sensing matrix. Furthermore, to obtain a concatenated model at $k$-th subcarrier, corresponding to all the $M$ blocks, we get
\begin{align}
    \underbrace{\begin{bmatrix}
        \mathbf{y}_{1}[k] \\ \vdots \\ \mathbf{y}_M[k] 
    \end{bmatrix}}_{\mathbf{y}[k]} \approx 
    \underbrace{\begin{bmatrix}
        \boldsymbol{\Psi}_{\mathcal{U},1}[k] \\ \vdots \\ \boldsymbol{\Psi}_{\mathcal{U},M}[k]
    \end{bmatrix}}_{\boldsymbol{\Psi}_{\mathcal{U}}[k]}\mathbf{h}_{b,\mathcal{U}}[k] + 
    \underbrace{\begin{bmatrix}
        \boldsymbol{\eta}_1[k] \\ \vdots \\ \boldsymbol{\eta}_M[k]
    \end{bmatrix}}_{\boldsymbol{\eta}[k]},
\end{align}
where $\mathbf{y}[k] \in \mathbb{C}^{M N_s \times 1}$ denotes the stacked pilot output, $\boldsymbol{\Psi}_{\mathcal{U}}[k] \in \mathbb{C}^{MN_s \times G_RG_T}$ represents stacked equivalent sensing matrix, $\boldsymbol{\eta}[k] \in \mathbb{C}^{MN_s \times 1}$ denotes stacked noise for all the $M$ blocks at $k$-th subcarrier. Moreover, the stacked noise covariance matrix $\mathbf{B} = \mathbb{E}\{\boldsymbol{\eta}[k]\boldsymbol{\eta}^H[k]\} = \mathrm{blkdiag}(\mathbf{R}_{\eta\eta,1}, \mathbf{R}_{\eta\eta,2}, \cdots \mathbf{R}_{\eta\eta,M}) \in \mathbb{C}^{MN_s \times MN_s}$. Note that the directional propagation characteristics in the THz band inherently result in an angularly sparse multipath channel \cite{mishra2017sparse}. Consequently, sparse reconstruction paradigms are particularly well-suited and can achieve excellent performance. It should be noted that combining the effects of beam-squint, frequency selectivity, and low-resolution ADCs in a THz system introduces \textit{severe nonlinearities}, as each impairment interacts jointly rather than additively. These distortions break the conventional linear assumptions used in sparse estimation, since the quantization process alters the signal statistics and the beam-squint effect induces subcarrier-dependent angular variations. Consequently, even advanced inference techniques encounter difficulties in forming a cohesive channel estimate across different frequencies and antennas. In this regard, variational Bayesian inference effectively mitigates these challenges by approximating the intractable joint posterior through a tractable variational distribution, enabling reliable channel recovery under joint quantization and beam-squint impairments. Therefore, the next section details a variational Bayesian inference-based approach for CSI estimation.
\vspace{-2mm}
\section{Variational Bayesian Learning based channel estimation}
\vspace{-1mm}
Bayesian probabilistic models are learning frameworks that assume unknown variables follow a posterior distribution. This posterior is formed by combining prior distributions, which represent beliefs about the variables before observing data, with a likelihood function that models the observed data. In this regard, let a parameterized Gaussian prior $f(\mathbf{h}_{b,\mathcal{U}}[k];\boldsymbol{\Gamma}_{k,\mathcal{U}})$ corresponding to the $k$-th subcarrier can be defined as
\vspace{-2mm}
\begin{align}
    f(\mathbf{h}_{b,\mathcal{U}}[k];\boldsymbol{\Gamma}_{k,\mathcal{U}}) = \prod_{t=1}^{G_RG_T}(\pi \gamma^{-1}_{k,t})^{-1} \mathrm{exp}\Big(-\frac{|\mathbf{h}_{b,\mathcal{U}}[k](t)|^2}{\gamma^{-1}_{k,t}}\Big), \label{parameterized}
\end{align}
where $\gamma_{k,t}$ represents the $t$-th hyperparameter corresponding to $k$-th subcarrier and $\boldsymbol{\Gamma}_{k,\mathcal{U}} = \mathrm{diag}(\gamma_{k,1}, \cdots, \gamma_{k,G_RG_T}) \in \mathbb{R}^{G_RG_T \times G_RG_T}$ represents the hyperparameter matrix.\\
\underline{Sparsity-enhancing Prior:} To promote sparsity in the estimation process, we adopt a hierarchical \textit{Gaussian-Gamma} prior framework \cite{tzikas2008variational}, which is widely used in machine learning. In this framework, a Gamma prior is imposed on the corresponding precision hyperparameters due to its conjugacy. While stationary Gaussian priors simplify Bayesian inference analytically, they are limited in their ability to model the distinct channel characteristics observed at each subcarrier. This limitation arises due to the spatial-wideband effect. Therefore, a nonstationary Gaussian prior with separate inverse variance parameters per subcarrier becomes essential to effectively model such local signal behaviors. In this regard, we constrain the hyperparameter $\boldsymbol{\bar{\gamma}}_k = ({\gamma}_{k,1},\cdots,{\gamma}_{k,G_RG_T})^T$ by treating them as random variables and imposing a Gamma prior on them as
\vspace{-2mm}
\begin{align}
    f(\boldsymbol{\bar{\gamma}}_{k};\mathsf{m},\mathsf{c}) &=  \mathrm{Gamma}(\gamma_{k,t}|\mathsf{m}_t,\mathsf{c}_t) \notag \\
    & = \frac{\mathsf{c}^{\mathsf{m}}}{\Upsilon(\mathsf{m})}\gamma_{k,t}^{\mathsf{m}-1}\mathrm{exp}(-\mathsf{c}\gamma_{k,t}), \; \gamma_{k,t} > 0, \label{standard_gamma}
\end{align}
where $\mathsf{m}>0$ and $\mathsf{c}>0$ controls the sparsity and sets the overall scale of the precision, respectively, while $\Upsilon(.)$ denotes the normalizing constant defined as $\Upsilon(\mathsf{m}) = \int_0^\infty u^{\mathsf{m}-1} e^{-u }du$. Therefore, the joint distribution over all the hyperparameters can be given as $f(\boldsymbol{\bar{\gamma}}_k) = \prod_{t=1}^{G_RG_T} f({\gamma}_{k,t}|\mathsf{m}_t,\mathsf{c}_t)$.\\
\underline{Modeling unknown noise precision:} Furthermore, the noise precision is positive and unknown quantity, hence, we impose Gamma prior distribution as
\vspace{-2mm}
\begin{align}
    f(\flat;\mathsf{w},\mathsf{e}) = \mathrm{Gamma}(\flat|\mathsf{w},\mathsf{e}),
\end{align}
where $\flat = \frac{1}{\sigma^2}$ denotes the noise precision. Note that, the high-dimensional integral $f(\mathbf{y}[k]) = \int f(\mathbf{y}[k],\mathcal{J}) d\mathcal{J}$ is mathematically intractable, where $\mathcal{J} = \{\mathbf{h}_{b,\mathcal{U}}[k], \boldsymbol{\bar{\gamma}}_k,\flat\}$ represents the set of latent variables. To address this, we employ variational Bayesian inference, which approximates the true posterior distribution $f(\mathcal{J}|\mathbf{y}[k])$ with a tractable variational distribution $g(\mathcal{J})$ that can be formulated as
\vspace{-5mm}

\small
\begin{align}
     \mathrm{ln}f(\mathbf{y}[k]) = \underbrace{\int g(\mathcal{J}) \mathrm{ln} \frac{f(\mathbf{y}[k],\mathcal{J})}{g(\mathcal{J})} d\mathcal{J}}_{\mathcal{L}(g)} - \underbrace{\int g(\mathcal{J}) \; \mathrm{ln} \frac{f(\mathcal{J}|\mathbf{y}[k])}{g(\mathcal{J})} d\mathcal{J}}_{\mathrm{KL}(g(\mathcal{J}) \parallel f(\mathcal{J}|\mathbf{y}[k]))}, \label{KL}
\end{align}
\normalsize
where $f(\mathbf{y}[k])$ denotes the marginal likelihood, $\mathcal{L}(g)$ represents the lower bound, and $\mathrm{KL}(g \parallel f)$ is the Kullback-Leibler divergence between the variational distribution $g(\mathcal{J})$ and true posterior $f(\mathcal{J}|\mathbf{y}[k])$, such that $\mathrm{KL}(g \parallel f) \geq 0$. Furthermore, Bayesian inference involves computing the posterior distribution which can be expressed as
\vspace{-5mm}

\small
\begin{align}
    f\big(\mathbf{h}_{b,\mathcal{U}}[k],\boldsymbol{\bar{\gamma}}_k, \flat|\mathbf{y}[k]\big) = \frac{f(\mathbf{y}[k]|\mathbf{h}_{b,\mathcal{U}}[k],\flat)f(\mathbf{h}_{b,\mathcal{U}}[k]|\boldsymbol{\bar{\gamma}}_k)f(\boldsymbol{\bar{\gamma}}_k)f(\flat)}{f(\mathbf{y}[k])}.
\end{align}
\normalsize
\begin{figure*}
    \begin{align}
     f(\mathbf{h}_{b,\mathcal{U}}[k],\boldsymbol{\bar{\gamma}}_k, \flat|\mathbf{y}[k];\mathsf{m},\mathsf{c},\mathsf{w},\mathsf{e}) \approx \prod_{k=1}^K g(\mathbf{h}_{b,\mathcal{U}}[k],\boldsymbol{\bar{\gamma}}_k, \flat) = \prod_{k=1}^K g(\mathbf{h}_{b,\mathcal{U}}[k]) g(\boldsymbol{\bar{\gamma}}_k) g(\flat),  \label{mean-field}
\end{align}
\hrulefill \vspace{-1.7\baselineskip}
\end{figure*}
\begin{figure*}
    \begin{align}
   \mathrm{ln} \, g(\mathbf{h}_{b,\mathcal{U}}[k]) &= \mathbb{E}_{ \boldsymbol{\bar{\gamma}}_k, \flat}\big\{ \mathrm{ln} \, f(\mathbf{y}[k]|\mathbf{h}_{b,\mathcal{U}}[k],\flat) + \mathrm{ln} \, f(\mathbf{h}_{b,\mathcal{U}[k]}|\boldsymbol{\bar{\gamma}}_k) + \mathrm{ln} \, f(\boldsymbol{\bar{\gamma}}_k) + \mathrm{ln} \,  f(\flat) \big\} + \mathrm{constant}, \label{h_estimate} \\
    \mathrm{ln} \, g(\mathbf{h}_{b,\mathcal{U}}[k]) &= \mathbb{E}_{ \boldsymbol{\bar{\gamma}}_k, \flat}\big\{\flat\big(\mathbf{y}[k]-\boldsymbol{\Psi}_{\mathcal{U}}[k]\mathbf{h}_{b,\mathcal{U}}[k]\big)^H \mathbf{B}^{-1}\big(\mathbf{y}[k]-\boldsymbol{\Psi}_{\mathcal{U}}[k]\mathbf{h}_{b,\mathcal{U}}[k]\big) - \frac{1}{2} \mathrm{tr}(\mathbf{\Gamma}_{k,\mathcal{U}}\mathbf{h}_{b,\mathcal{U}}[k]\mathbf{h}_{b,\mathcal{U}}^H[k]) \big\} + \mathrm{constant}. \label{h_solve}
\end{align}
\hrulefill \vspace{-1.6\baselineskip}
\end{figure*}
\begin{figure*}
\begin{align}
    \boldsymbol{\mathfrak{Q}}[k] &= \big(\mathbb{E}\{\mathbf{\Gamma}_{k,\mathcal{U}}\}\big)^{-1} - \big(\mathbb{E}\{\mathbf{\Gamma}_{k,\mathcal{U}}\}\big)^{-1} \mathbb{E}\{\flat\}\mathbf{\Psi}_{\mathcal{U}}^H[k]\Big(\mathbf{B} + \mathbf{\Psi}_{\mathcal{U}}[k]\big(\mathbb{E}\{\mathbf{\Gamma}_{k,\mathcal{U}}\}\big)^{-1}\mathbb{E}\{\flat\}\mathbf{\Psi}_{\mathcal{U}}^H[k]\Big)^{-1}\mathbf{\Psi}_{\mathcal{U}}[k]\big(\mathbb{E}\{\mathbf{\Gamma}_{k,\mathcal{U}}\}\big)^{-1}, \label{wood_covariance} \\
    \boldsymbol{\mu}[k] &= \mathbb{E}\{\flat\}\big(\mathbb{E} \big\{\mathbf{\Gamma}_{k,\mathcal{U}}\})^{-1} \mathbf{\Psi}_{\mathcal{U}}^H[k]\mathbf{B}^{-1}\Big(\mathbf{I}_{MN_{\mathrm{RF}}^R} + \mathbb{E}\{\flat\}\mathbf{\Psi}_{\mathcal{U}}[k]\big(\mathbb{E}\{\mathbf{\Gamma}_{k,\mathcal{U}}\}\big)^{-1}\mathbf{\Psi}_{\mathcal{U}}^H[k]\mathbf{B}^{-1}\Big)^{-1}\mathbf{y}[k], \label{wood_mean}
\end{align}
\hrulefill \vspace{-1.7\baselineskip}
\end{figure*}
\begin{figure*}
    \begin{align}
        \mathrm{ln} \, \{g(\boldsymbol{\bar{\gamma}}_k)\} & = \mathbb{E}_{\mathbf{h}_{b,\mathcal{U}}[k]} \{\mathrm{ln} \, f(\mathbf{h}_{b,\mathcal{U}}[k]|\boldsymbol{\bar{\gamma}}_k) + \mathrm{ln} \, f(\boldsymbol{\bar{\gamma}}_k) \} + \mathrm{constant}, \label{post_h} \\
        \mathrm{ln} \, \{g(\boldsymbol{\bar{\gamma}}_k)\} & = -\sum_{t=1}^{G_R G_T} \mathrm{ln}(\gamma_{k,t}) - \sum_{t=1}^{G_RG_T} \gamma_{k,t} \mathbb{E}\big\{\mathbf{h}_{b,\mathcal{U}}^H[k](t)\mathbf{h}_{b,\mathcal{U}}[k](t)\big\} + (\mathsf{m}+1) \sum_{t=1}^{G_RG_T} \mathrm{ln} (\gamma_{k,t}) - \mathsf{c}\sum_{t=1}^{G_RG_T} \gamma_{k,t} + \mathrm{cons}, \label{posth} \\
        \mathrm{ln} \, \{g(\boldsymbol{\bar{\gamma}}_k)\} & = (\mathsf{m}+1-1) \sum_{t=1}^{G_RG_T} \mathrm{ln}(\gamma_{k,t}) - \sum_{t=1}^{G_RG_T} \Big(\mathsf{c} + \mathbb{E}\big\{\mathbf{h}_{b,\mathcal{U}}^H[k](t)\mathbf{h}_{b,\mathcal{U}}[k](t)\big\}\Big)\gamma_{k,t} + \mathrm{constant} \label{posth_final}.
    \end{align} 
    \hrulefill \vspace{-1.7\baselineskip}
\end{figure*}
\hspace{-2.5mm}Since, the marginal likelihood is intractable, as a result, the posterior normalization constant cannot be evaluated in closed form. Therefore, we adopt the \textit{mean-field} approximation, which assumes that the components of the latent variable to be mutually independent such that $(\cup_{s=1}^S \mathcal{J}_s = \mathcal{J} $ and $\mathcal{J}_s \cap \mathcal{J}_j = \varnothing)$, which is further given in Eq. \eqref{mean-field}. 
Substituting Eq. \eqref{mean-field} in Eq. \eqref{KL} and maximizing $\mathcal{L}(g)$ \cite{bishop2006pattern}, one obtain
\begin{align}
    g(\mathcal{J}_s)  \varpropto \mathrm{exp}\big\{\mathbb{E}_{ \thicksim g(\mathcal{J}_s)}(\mathrm{ln} \; f(\mathbf{y}[k], \mathcal{J})) \big\},
\end{align}
where $\mathbb{E}_{\thicksim g(\mathcal{J}_s)}$ represents the expectation with respect to all the other parameters except $g(\mathcal{J}_s)$. The posterior distribution calculation of $g(\mathbf{h}_{b,\mathcal{U}}[k]), g(\boldsymbol{\bar{\gamma}}_k), g(\flat)$ are detailed below.\\
\underline{Computation of $g(\mathbf{h}_{b,\mathcal{U}}[k])$, \textbf{for} $k=1$ to $K$:} Keeping the terms that depends on $\mathbf{h}_{b,\mathcal{U}}[k]$, the joint posterior can be expressed as $g(\mathbf{h}_{b,\mathcal{U}}[k]) = \mathbb{E}\big\{f(\mathbf{y}[k], \mathbf{h}_{b,\mathcal{U}}[k],\boldsymbol{\bar{\gamma}}_k,\flat)_{g(\boldsymbol{\bar{\gamma}}_k),g(\flat)}\big\}$. Furthermore, after taking the log on both sides of the joint posterior, we obtain Eq. \eqref{h_estimate} which can be simplified to Eq. \eqref{h_solve}. One can observe that the resulting function is quadratic and identical with the standard Gaussian distribution $g(\mathbf{h}_{b,\mathcal{U}}[k]) \approx \mathcal{CN}(\boldsymbol{\mu}[k], \boldsymbol{\mathfrak{Q}}[k])$, which yields
\begin{align}
    \boldsymbol{\mathfrak{Q}}[k] &= \big(\mathbb{E}\{\mathbf{\Gamma}_{k,\mathcal{U}}\}+\mathbb{E}\{\flat\}\mathbf{\Psi}^H_{\mathcal{U}}[k]\mathbf{B}^{-1}\mathbf{\Psi}_{\mathcal{U}}[k]\big)^{-1}, \notag \\
    \boldsymbol{\mu}[k] &= \mathbb{E}\{\flat\}\boldsymbol{\mathfrak{Q}}[k] \boldsymbol{\Psi}_{\mathcal{U}}^H[k]\mathbf{B}^{-1}\mathbf{y}[k].
\end{align}
{\underline{Computation of $g(\boldsymbol{\bar{\gamma}}_k)$:}} Similarly, taking the terms that depend on $\boldsymbol{\bar{\gamma}}_k$ of the joint posterior and applying the $\mathrm{log}(\cdot)$ function on both sides of the equation, we obtain Eq. \eqref{post_h}. Proceeding with the algebraic simplification as shown in Eq. \eqref{posth}-\eqref{posth_final} and comparing with the standard Gamma distribution expressed in Eq. \eqref{standard_gamma}, we obtain
\vspace{-2mm}
\begin{align}
    \tilde{\mathsf{m}} &= \mathsf{m} + 1, \notag \\
    \tilde{\mathsf{c}} &= \mathsf{c} + \boldsymbol{\mu}^H[k](t) \boldsymbol{\mu}[k](t) + \boldsymbol{\mathfrak{Q}}[k](t,t). \label{hyperparameter-update}
\end{align}
Moreover, the quantity $\mathbb{E}_{g(\boldsymbol{\bar{\gamma}}_k)}(\mathbf{\Gamma}_{k,\mathcal{U}}) = \mathrm{diag}\big([\mathbb{E}_{g(\boldsymbol{\bar{\gamma}}_k)}(\gamma_1), \cdots, \mathbb{E}_{g(\boldsymbol{\bar{\gamma}}_k)}(\gamma_{G_R G_T})] \big)$ where $\mathbb{E}_{g(\bar{\boldsymbol{\gamma}}_k)}(\gamma_{k,t}) = \frac{\tilde{\mathsf{m}}}{\tilde{\mathsf{c}}}$ and $\mathsf{m}, \mathsf{c}$ are set as described in Section-\ref{simulation} to induce a non-informative prior on $\gamma_{k,t}$.\\
\begin{figure*}[t]
    \begin{align}
        \mathrm{ln} (g(\flat)) &= \mathbb{E}_{\mathbf{h}_{b,\mathcal{U}}[k]}\big\{\mathrm{ln} \, f(\mathbf{y}[k]|\mathbf{h}_{b,\mathcal{U}}[k],\flat) + \mathrm{ln} \, f(\flat)\big\} + \mathrm{constant}, \label{precision}\\
        \mathrm{ln} (g(\flat)) &= \mathbb{E}_{\mathbf{h}_{b,\mathcal{U}}[k]}\big\{MN_{\mathrm{RF}}^R \mathrm{ln}(\flat) - \flat\big(\mathbf{y}[k]-\boldsymbol{\Psi}_{\mathcal{U}}[k]\mathbf{h}_{b,\mathcal{U}}[k]\big)^H \mathbf{B}^{-1}\big(\mathbf{y}[k]-\boldsymbol{\Psi}_{\mathcal{U}}[k]\mathbf{h}_{b,\mathcal{U}}[k]\big) + (\mathsf{w}-1) \mathrm{ln}(\flat) + \mathsf{e}\flat \big\} + \mathrm{constant},\label{precision_ini} \\
        \mathrm{ln} (g(\flat)) &= \mathbb{E}_{\mathbf{h}_{b,\mathcal{U}}[k]} \big\{(\mathsf{w}+MN_{\mathrm{RF}}^R -1) \mathrm{ln}(\flat) + \flat\big[\big(\mathbf{y}[k]-\boldsymbol{\Psi}_{\mathcal{U}}[k]\mathbf{h}_{b,\mathcal{U}}[k]\big)^H \mathbf{B}^{-1}\big(\mathbf{y}[k]-\boldsymbol{\Psi}_{\mathcal{U}}[k]\mathbf{h}_{b,\mathcal{U}}[k]\big) + \mathsf{e}\big] \big\} \label{precision_final}.
    \end{align}
    \hrulefill \vspace{-1.5\baselineskip}
\end{figure*}
\hspace{-2mm} \underline{Computation of $g(\flat)$:} To derive the update for $\flat$, we isolate the terms in the joint posterior that depend on $\flat$, and then apply the $\mathrm{log}(.)$ to both sides of the expression yielding Eq. \eqref{precision}. Similarly, by further simplifying Eq. \eqref{precision_ini}-\eqref{precision_final}, the updates pertaining to $\flat$ are obtained as
\begin{align}
     \mathsf{w} &= \mathsf{w} + MN_{\mathrm{RF}}^R, \notag \\
     \mathsf{e} &= \mathsf{e} + \mathbb{E}_{\mathbf{h}_{b,\mathcal{U}}[k]}\Big\{\big(\mathbf{y}[k]-\boldsymbol{\Psi}_{\mathcal{U}}[k]\mathbf{h}_{b,\mathcal{U}}[k]\big)^H \mathbf{B}^{-1} \notag \\ & \qquad\qquad\qquad\qquad \big(\mathbf{y}[k]-\boldsymbol{\Psi}_{\mathcal{U}}[k]\mathbf{h}_{b,\mathcal{U}}[k]\big)\Big\}. \label{noise_update}
\end{align}
Note that the resulting variational inference algorithm not only provides closed-form updates for each variable, but also accounts for the uncertainties associated with the other estimated parameters. Algorithm-\ref{est_algo} provides an overview of the Bayesian inference procedure. The algorithm repeats until the convergence condition is satisfied. Furthermore, the final estimate $\widehat{\mathbf{h}}_{\mathrm{BI}}[k] \in \mathbb{C}^{G_RG_T \times 1}$ is expressed as
\begin{align}
    \widehat{\mathbf{H}}_{\mathrm{BI}}[k] = \mathbf{A}_R(\Theta_R,f_k)\mathrm{vec}^{-1}(\widehat{\mathbf{h}}_{\mathrm{BI}}[k])\mathbf{A}_{T,\mathcal{U}}^H({\Phi}_{T},f_k),
\end{align}
\sloppy
\begin{figure*}
    \begin{align}
    \mathrm{ln} \, f(\mathbf{h}_{b,\mathcal{U}}; \mathbf{\Gamma}_{k,\mathcal{U}})& = -\flat\big(\mathbf{y}[k]-\boldsymbol{\Psi}_{\mathcal{U}}[k]\mathbf{h}_{b,\mathcal{U}}[k]\big)^H \mathbf{B}^{-1}\big(\mathbf{y}[k]-\boldsymbol{\Psi}_{\mathcal{U}}[k]\mathbf{h}_{b,\mathcal{U}}[k]\big) + MN_{\mathrm{RF}}^R \mathrm{ln} \, \flat - MN_{\mathrm{RF}}^R \mathrm{ln} \, \pi -  \sum_{t=1}^{G_RG_T}\big[\mathrm{ln}(\pi \gamma_{k,t}) + \notag \\ & \frac{1}{2} \mathrm{tr}(\mathbf{\Gamma}_{k,\mathcal{U}}\mathbf{h}_{b,\mathcal{U}}[k]\mathbf{h}_{b,\mathcal{U}}^H[k])\big] + (\mathsf{m}-1) \sum_{t=1}^{G_RG_T} \mathrm{ln} (\gamma_{k,t}) - \mathsf{c} \sum_{t=1}^{G_RG_T} \gamma_{k,t} + (\mathsf{w}-1) \mathrm{ln} \, \flat - \mathsf{e} \, \flat + \mathrm{constant}. \label{complete_log}
\end{align}
\hrulefill \vspace{-1\baselineskip}
\end{figure*}
where $\mathbf{A}_{T,\mathcal{U}}({\Phi}_{T},f_k) = \mathrm{blkdiag}(\mathbf{A}_{T,1}(\Phi_{T,1},f_k), \cdots, \\ \mathbf{A}_{T,U}(\Phi_{T,U},f_k)) \in \mathbb{C}^{N_TG_T \times N_TG_T}$. 
\begin{algorithm}[t]
\DontPrintSemicolon 
\textbf{Input}: Received signal $\{\mathbf{y}[k]\}_{k=1}^K,$ equivalent sensing matrix $\{\mathbf{\Psi}_{\mathcal{U}}[k]\}_{k=1}^K$, noise covariance $\mathbf{B}^{-1}$, constants of gamma distribution $\mathsf{m},\mathsf{c},\mathsf{w},\mathsf{e}$ and $\varepsilon, N_\mathrm{max}$

\textbf{Initialization:} $\gamma_{k,t}^{(0)} = 1 \, \forall \, 0 \, \leq t < G_RG_T; \widehat{\mathbf{\Gamma}}_{k,\mathcal{U}}^{(0)} = \mathbf{I}_{G_RG_T}; \widehat{\mathbf{\Gamma}}_{k,\mathcal{U}}^{(-1)} = \mathbf{0}; \iota=0$
 
\While{$\parallel \widehat{\mathbf{\Gamma}}_{k,\mathcal{U}}^{(\iota)} - \widehat{\mathbf{\Gamma}}_{k,\mathcal{U}}^{(\iota-1)} \parallel_{\mathcal{F}}^2 \geq \varepsilon \; \mathrm{and} \; \iota < N_{\mathrm{max}}$}
{
$\iota = \iota+1$

Update the \textit{a posteriori} covariance and mean using Eq. \eqref{wood_covariance} and \eqref{wood_mean}, respectively.

Update the hyperparameters and constants of Gamma distribution using Eq. \eqref{hyperparameter-update}.

Update the noise covariance and constants of Gamma distribution using Eq. \eqref{noise_update}.
}

\textbf{Output:~~}{$\widehat{\mathbf{h}}_{\mathrm{BI}}[k] = \boldsymbol{\mu}[k]$}
\caption{Variational Bayesian inference for sparse dual-wideband channel estimation} \vspace{-0.2 \baselineskip}
\label{est_algo}
\end{algorithm}

As detailed in Section-\ref{contri}, designing the hybrid beamformer with TTD elements requires the optimal beam-steering angles for each RF chain. These angles configure the analog beamformer so that the TTD elements can introduce the requisite frequency-dependent delays, thereby counteracting the beam-squint effects that arise in the wideband THz operation. Moreover, leveraging the angular information that naturally emerges in the Bayesian estimated beamspace channel offers a simple yet highly efficient alternative to conventional beam training or exhaustive codebook searches \cite{he2017codebook}. In this regard, the variational Bayesian inference intrinsically yields an improved sparse channel estimate in which the non-zero coefficients directly correspond to the \textit{dominant propagation paths}, thereby implicitly encoding reliable steering direction information. Thus, by prioritizing these posterior magnitudes and mapping their indices to the corresponding AoA/AoD pairs, the required steering angles can be effectively obtained without additional pilot overhead, iterative search procedures, or dictionary adjustments. This significantly reduces computational complexity and enables seamless integration of the channel training procedure with analog beamformer configuration. 

Algorithm-\ref{opt_ang} summarizes the procedure for extracting the AoA/AoD pairs directly from the estimated beamspace channel. Step-$3$ sorts the estimated hyperparameters for each user in decreasing order. Step-$4$ identifies the $N_{\mathrm{RF}}$ dominant AoD indices, while Step-$5$ determines the AoA indices by applying a modulo operation with respect to the receive dictionary size. Step-$6$ appends the AoD and AoA indices for each user and returns the dominant transmit and receive angular indices. As detailed earlier, the proposed algorithm utilizes only partial CSI information, specifically, the indices corresponding to the quantized AoA/AoD pairs. Consequently, it circumvents the need to store the complete CSI at the receiver or to fully feed it back to the transmitter, making it low-complex and very well suited for practical implementations. The next section will discuss the convergence of the algorithm.
\begin{algorithm}[t] \label{opt_ang}
\DontPrintSemicolon 
\textbf{Input:} Estimated beamspace channel $\{\widehat{\mathbf{h}}_{\mathrm{BI}}[k]\}_{k=1}^K$, transmit RF chain $N_{\mathrm{RF},u}^T$, receive RF chain $N_{\mathrm{RF}}^R$

\textbf{Initialization:} $\tilde{\Phi}_T[k] = [\,\,], \tilde{\Theta}_R[k] = [\,\,]$
 
\For{$k=1:K$}{
$\tilde{\Phi}_T[k] = [\,\,]; \tilde{\Theta}_R[k] = [\,\,]$

\For{$u = 1:U$}
{
$\widehat{\mathbf{h}}_u = \widehat{\mathbf{h}}_{\mathrm{BI}}[k][:,(u-1)G_{T,u}+1:uG_{T,u}]$

$[\sim,\mathrm{I}] = \mathrm{sort}(|\widehat{\mathbf{h}}_u|,\mathrm{descend})$\\
  \For{$\mathsf{n}=1:N_{\mathrm{RF},u}$}
  {
  $\grave{\Phi}_{T,u}[\mathsf{n}] = \frac{\lfloor \mathrm{I}(\mathsf{n})-1 \rfloor}{G_{T,u}}+1$
  }
  \For{$\mathsf{n}=1:\frac{N^R_{\mathrm{RF}}}{U}$}
  {
  $\grave{\Theta}_{R,u}[\mathsf{n}] = \mathrm{rem}(\mathrm{I}(\mathsf{n})-1,G_R^{l})+1$
  }
  
   $\tilde{\Phi}_T[k] = [\tilde{\Phi}_T[k] \quad \grave{\Phi}_{T,u}]$, $\tilde{\Theta}_R[k] = [\tilde{\Theta}_R[k] \quad \grave{\Theta}_{R,u}]$
}
}
    
\textbf{Output:~~}{$\tilde{\Phi}_T[k], \tilde{\Theta}_R[k]$}
\caption{Determination of optimal physical angles from the estimated beamspace}
\end{algorithm}
\vspace{-3mm}
{ \subsection{Convergence Analysis}\label{convergence_VB}
The proposed variational Bayesian inference algorithm is formulated as a coordinate ascent procedure that iteratively maximizes the evidence lower bound (ELBO), $\mathcal{L}(g)$, defined as
\vspace{-2mm}
\begin{equation}
    \mathcal{L}(g)=\int g(\mathcal{J})\ln{\frac{f(\mathbf{y}[k],\mathcal{J})}{g(\mathcal{J})}}d\mathcal{J},
\end{equation}
which satisfies $\mathcal{L}(g) \leq \ln f(\mathbf{y}[k])$ since $\mathrm{KL}(g(\mathcal{J}) \parallel f(\mathcal{J}|\mathbf{y}[k])) \geq 0$ \cite{bishop2006pattern}. Under the \textit{mean-field} factorization $g(\mathcal{J})=g(\mathbf{h}_{b,\mathcal{U}}[k])\cdot g(\bar{\boldsymbol{\gamma}}_k)\cdot g(\flat)$, each coordinate update takes the form
\vspace{-2mm}
\begin{equation}
    \ln{g(\mathcal{J}_s)}=\mathbb{E}_{\sim g(\mathcal{J}_s)}\left[\ln{f(\mathbf{y}[k],\mathcal{J})}\right]+\mathrm{const},
\end{equation}
which is equivalent to minimizing the KL divergence between the variable distribution and the corresponding optimal target distribution $\tilde{f}(\cdot)$, i.e., $\mathcal{L}(g(\mathcal{J}_s))=-\mathrm{KL}(g(\mathcal{J}_s)\parallel\tilde{f})+\mathrm{const}$, where $\tilde{f}$ represents the effective distribution for one variable. For $\mathbf{h}_{b,\mathcal{U}}[k]$ it is defined as 
\vspace{-2mm}
\begin{equation}
    \ln \tilde{f}(\mathbf{y}[k],\mathbf{h}_{b,\mathcal{U}}[k])=\mathbb{E}_{g(\bar{\gamma}_k),g(\flat)}\left[\ln f(\mathbf{y}[k],\mathcal{J})\right].
\end{equation}
Additionally KL divergence is convex in its first argument, which means each individual coordinate update finds the unique global optimum, guaranteeing $\mathcal{L}^{(\iota+1)}\geq\mathcal{L}^{(\iota)}$ at every iteration $\iota$ \cite{nguyen2024variational}. Given that the ELBO is bounded by the finite log marginal likelihood $\ln f(\mathbf{y}[k]) < \infty$, which holds due to bounded transmit power, finite noise variance, and the proper \textit{Gaussian-Gamma} hierarchical prior. This monotone convergence theorem guarantees convergence of the ELBO sequence to a finite limit $\mathcal{L}^* < \infty$, where $\mathcal{L}^*$ represents optimal ELBO for each posterior i.e., $\mathbf{h}_{b,\mathcal{U}}[k], \bar{\boldsymbol{\gamma}}_k, \flat$.} The next section will describe the MU BCRB.
\vspace{-4mm}
\subsection{MU Bayesian Cram{\'e}r-Rao lower bound}
\vspace{-1mm}
From the parameterized Gaussian prior defined in Eq. \eqref{parameterized}, the complete log-likelihood of the beamspace channel $\mathbf{h}_{b,\mathcal{U}}[k]$ can be expressed as Eq. \eqref{complete_log}. Moreover, by taking the log-likelihood of the conditional probability density function (PDF), we obtain
\begin{align}
    f(&\mathbf{y}[k] |\mathbf{h}_{b,\mathcal{U}}[k],\flat) = -\flat\big(\mathbf{y}[k]-\boldsymbol{\Psi}_{\mathcal{U}}[k]\mathbf{h}_{b,\mathcal{U}}[k]\big)^H \mathbf{B}^{-1} \notag \\ & \big(\mathbf{y}[k]-\boldsymbol{\Psi}_{\mathcal{U}}[k]\mathbf{h}_{b,\mathcal{U}}[k]\big) + MN_{\mathrm{RF}}^R \mathrm{ln} \, \flat - MN_{\mathrm{RF}}^R \mathrm{ln} \pi.
\end{align}
Let the Bayesian Fisher information matrix (FIM) $\mathbf{T}[k] \in \mathbb{C}^{G_RG_T \times G_RG_T}$, be obtained as $\mathbf{T}[k] = \mathbf{T}_p[k] + \mathbf{T}_b[k],$ where $\mathbf{T}_p[k] \in \mathbb{C}^{G_RG_T \times G_RG_T}$ denotes FIM corresponding to pilot output $\mathbf{y}[k]$ while $\mathbf{T}_b[k] \in \mathbb{C}^{G_RG_T \times G_RG_T}$ represents the FIM corresponding to \textit{a prior} information of the beamspace domain $\mathbf{h}_{b,\mathcal{U}}[k]$. Let $\mathrm{ln}(.)$ be represented as $\mathcal{L}(.)$ for notational simplicity. Therefore, one can mathematically represent the FIM as
\vspace{-2mm}
\begin{align}
    \mathbf{T}_p[k] &= -\mathbb{E}_{\mathbf{y}[k],\mathbf{h}_{b,\mathcal{U}}[k]}\bigg\{\frac{\partial^2 \mathcal{L}(\mathbf{y}[k]|\mathbf{h}_{b,\mathcal{U}}[k],\flat)}{\partial \mathbf{h}_{b,\mathcal{U}}[k] \mathbf{h}_{b,\mathcal{U}}^H[k]} \bigg\}, \\
    \mathbf{T}_b[k] &= - \mathbb{E}_{\mathbf{h}_{b,\mathcal{U}}[k]}\bigg\{\frac{\partial^2\mathcal{L}(\mathbf{h}_{b,\mathcal{U}}[k];\mathbf{\Gamma}_{k,\mathcal{U}})}{\partial \mathbf{h}_{b,\mathcal{U}}[k] \mathbf{h}_{b,\mathcal{U}}^H[k]}\bigg\}.
\end{align}
Substituting the corresponding expressions for the conditional PDF and the parameterized prior, the closed-form expression of the FIM can be obtained as
\vspace{-2mm}
\begin{align}
    \mathbf{T}_p[k] = \flat\mathbf{\Psi}_{\mathcal{U}}^H[k]\mathbf{B}^{-1}\mathbf{\Psi}_{\mathcal{U}}[k], \mathbf{T}_b[k] = \widehat{\boldsymbol{\Gamma}}_{k,\mathcal{U}}.
\end{align}
Therefore, the BCRLB provides a fundamental lower bound on the mean squared error (MSE) of any unbiased channel estimator, and is expressed as the inverse of the Bayesian FIM, given as
\vspace{-2mm}
\begin{align}
    \mathrm{MSE}(\widehat{\mathbf{h}}_{\mathrm{BI}}[k]) \geq \mathrm{Tr}(\mathbf{\Psi}_{\mathcal{U}}[k]\mathbf{T}^{-1}[k]\mathbf{\Psi}^H_{\mathcal{U}}[k]),
\end{align}
where $\mathrm{MSE}(.)$ is defined as $\parallel \widehat{\mathbf{h}}_{\mathrm{BI}}[k]-\mathbf{h}_{b,\mathcal{U}}[k]\parallel^2_2$ for each subcarrier. The next section introduces the design framework for a TTD-based hybrid transceiver architecture.
\vspace{-3mm}
\subsection{Computational Complexity}
\vspace{-1mm}
{ This subsection presents the computational complexity analysis of the proposed Bayesian inference framework and the existing sparse recovery techniques, namely OMP \cite{karabulut2004sparse} and MFOCUSS \cite{cotter2005sparse}. Table-\ref{table111} summarizes the per-subcarrier computational complexity of the OMP algorithm, including the key computational steps and their corresponding FLOP counts. The overall computational complexity can be expressed as $KG_RG_TMN_s+K(\frac{2}{3}j^3+j^2MN_s+3jMN_s)$, where $j$ denotes the iteration index. The worst-case complexity is dominated by the LS estimation step, resulting in an order of $\mathcal{O}(KM^3N_s^3)$. The per-subcarrier computational complexity of the MFOCUSS algorithm is summarized in Table-\ref{table112}, including the main computational steps and their corresponding FLOP counts. The worst case computational complexity is of the order $\mathcal{O}(G_R^2G_T^2)$. It is noteworthy that the MFOCUSS algorithm incurs higher computational complexity compared to OMP. This is primarily due to the matrix inversion operations and iterative reweighting steps involved in solving the underlying optimization problem at each iteration. However, as demonstrated in the simulation results, MFOCUSS achieves improved estimation performance over OMP, owing to its ability to exploit joint sparsity and provide more accurate recovery of sparse channel coefficients through iterative refinement. Table-\ref{table113} presents the computational complexity of the Bayesian inference framework, along with the key computational steps and their associated FLOP counts. The worst-case complexity is on the order of $\mathcal{O}(G_R^3G_T^3),$ which is higher than that of OMP and MFOCUSS, since $G_R, G_T \gg M$. However, as demonstrated in the simulation results, the variational Bayesian approach achieves superior estimation performance compared to these techniques. This highlights a clear trade-off between computational complexity and estimation accuracy.}
\renewcommand{\arraystretch}{1}
\begin{table*}[t]
\centering
{ \caption{Computational complexity of the OMP scheme, per-subcarrier, in the $j$th iteration}
\vspace{-2mm}
\label{table111}
\resizebox{0.98\textwidth}{!}{%
\begin{tabular}{|c|c|c|c|}\hline
\textbf{Operation} & \textbf{Complex multiplications} & \textbf{Complex additions}& \textbf{FLOPs}\\ \hline \hline
$\boldsymbol{\Psi}^H_\mathcal{U}[k]\mathbf{T}(:,k)$ & $G_RG_TMN_s$ &$G_RG_T(MN_s-1)$  &$2G_RG_TMN_s-G_RG_T$\\\hline
$\hat{\mathbf{H}}_{\mathrm{LS},\mathcal{U}}(:,k)=(\boldsymbol{\Psi}^A_\mathcal{U})^{\dagger}\mathbf{Y}_{\mathcal{U}}(:,k)$ & $\frac{2}{3}j^3+j^2MN_s+2jMN_s$ &$\frac{2}{3}j^3+j^2(MN_s-1)+j(MN_s-1)$  &\makecell{$(G_RG_T)^3+(G_RG_T)^2+G_RG_T+$\\$2G_RG_TMN_sK-G_RG_TK$} \\\hline
$\mathbf{Y}_{\mathcal{U}}(:,k)-\boldsymbol{\Psi}^A_\mathcal{U}[k]\hat{\mathbf{H}}_{LS,\mathcal{U}}(:,k)$ & $jMN_s$ & $jMN_s$ &--\\\hline
\end{tabular}
}
\vspace{-1.52 \baselineskip}
}
\end{table*}
\renewcommand{\arraystretch}{1}
\begin{table*}[t]
\centering
{ \caption{Computational complexity of the MFOCUSS framework per sub-carrier}
\vspace{-2mm}
\label{table112}
\resizebox{0.98\textwidth}{!}{%
\begin{tabular}{|c|c|c|c|}\hline
\textbf{Operation} & \textbf{Complex multiplications} & \textbf{Complex additions} & \textbf{FLOPs} \\ \hline \hline
$\mathbf{\Psi}_{\mathcal{U}}[k]\mathbf{\Pi}^{-1}(\mathbf{h}_{b,\mathcal{U}}[k])\mathbf{\Psi}_{\mathcal{U}}^T[k]$ & $MN_s(G_RG_T)^2$ & $MN_sG_RG_T(G_RG_T-1)$&$2MN_s(G_RG_T)^2-MN_sG_RG_T$\\\hline
 $(\mathbf{\Psi}_{\mathcal{U}}[k]\mathbf{\Pi}^{-1}(\mathbf{h}_{b,\mathcal{U}}[k])\mathbf{\Psi}_{\mathcal{U}}^T[k])^{-1}$&$\frac{\left(MN_{s}\right)^3}{2} + \frac{3\left(MN_{s}\right)^2}{2}$  &  $\frac{\left(MN_{s}\right)^3}{2} - \frac{\left(MN_{s}\right)^2}{2}$&$(MN_s)^3+(MN_s)^2+MN_s $\\\hline
\end{tabular}
}
\vspace{-1.2 \baselineskip}
}
\end{table*}  
\renewcommand{\arraystretch}{1}
\begin{table*}[t]
\centering
{ \caption{Computational complexity of the Bayesian inference per-subcarrier, per-EM iteration}
\vspace{-2mm}
\label{table113}
\resizebox{0.98\textwidth}{!}{%
\begin{tabular}{|c|c|c|c|}\hline
\textbf{Operation} & \textbf{Complex multiplications} & \textbf{Complex additions} & \textbf{FLOPs} \\ \hline \hline
 $(\mathbb{E}({\mathbf\Gamma_{k,\mathcal{U}}}))^{-1}$&$\frac{\left(G_RG_T\right)^3}{2} + \frac{3\left(G_RG_T\right)^2}{2}$  &  $\frac{\left(G_RG_T\right)^3}{2} - \frac{\left(G_RG_T\right)^2}{2}$&$(G_RG_T)^3+(G_RG_T)^2+G_RG_T $\\\hline
 $\mathbf{\Psi}_{\mathcal{U}}[k](\mathbb{E}({\mathbf\Gamma_{k,\mathcal{U}}}))^{-1}\mathbf{\Psi}_{\mathcal{U}}^H[k]$& $MN_s(G_RG_T)^2$ & $MN_sG_RG_T(G_RG_T-1)$&$2MN_s(G_RG_T)^2-MN_sG_RG_T$\\\hline
 $(\mathbf{B}+\mathbf{\Psi}_{\mathcal{U}}[k](\mathbb{E}({\mathbf\Gamma_{k,\mathcal{U}}}))^{-1}\mathbf{\Psi}_{\mathcal{U}}^H[k])^{-1}$&$\frac{\left(MN_{s}\right)^3}{2} + \frac{3\left(MN_{s}\right)^2}{2}$  &  $\frac{\left(MN_{s}\right)^3}{2} - \frac{\left(MN_{s}\right)^2}{2}$&$(MN_s)^3+(MN_s)^2+MN_s $\\\hline
 $\mathbf{B}^{-1}$&$\frac{\left(MN_{s}\right)^3}{2} + \frac{3\left(MN_{s}\right)^2}{2}$  &$\frac{\left(MN_{s}\right)^3}{2} - \frac{\left(MN_{s}\right)^2}{2}$&$(MN_s)^3+(MN_s)^2+MN_s $\\\hline
 $(\mathbf{I}_{MN_{RF}^R}+\mathbf{\Psi}_{\mathcal{U}}[k](\mathbb{E}({\mathbf\Gamma_{k,\mathcal{U}}}))^{-1}\mathbf{\Psi}_{\mathcal{U}}^H[k]\mathbf{B}^{-1})^{-1}$&$\frac{\left(MN_{RF}^R\right)^3}{2} + \frac{3\left(MN_{RF}^R\right)^2}{2}$  &  $\frac{\left(MN_{RF}^R\right)^3}{2} - \frac{\left(MN_{RF}^R\right)^2}{2}$&$(MN_{RF}^R)^3+(MN_{RF}^R)^2+MN_{RF}^R $\\\hline
 $\boldsymbol{\mu}^H[k](t)\boldsymbol{\mu}[k](t)$ & $G_RG_T$& $G_RG_T-1$ &$2G_RG_T-1$\\\hline
\end{tabular}%
}
\vspace{-2 \baselineskip}
}
\end{table*}
\vspace{-3mm}
\section{TTD based Hybrid Transceiver design} \label{transceiver_design}
\vspace{-1mm}
In this section, we develop a beam-squint mitigation strategy in which frequency-dependent TTD elements perform coarse delay compensation across the signal bandwidth, followed by a bank of frequency-flat phase shifters (PS) that refine the residual steering error. Accordingly, the overall beamformer design is structured into two stages; the first stage involves identifying the optimal frequency-independent beam steering angles, while the second stage utilizes these angles to construct frequency-dependent RF precoders and combiners. Furthermore, to determine the angles in the first stage, we leverage the estimated beamspace AoA/AoD pairs obtained from Algorithm-\ref{opt_ang} to configure the TTD elements. The detailed procedure for converting the frequency-independent solution into a frequency-dependent is described next.
\begin{figure}
\centering
{\includegraphics[scale=0.18]{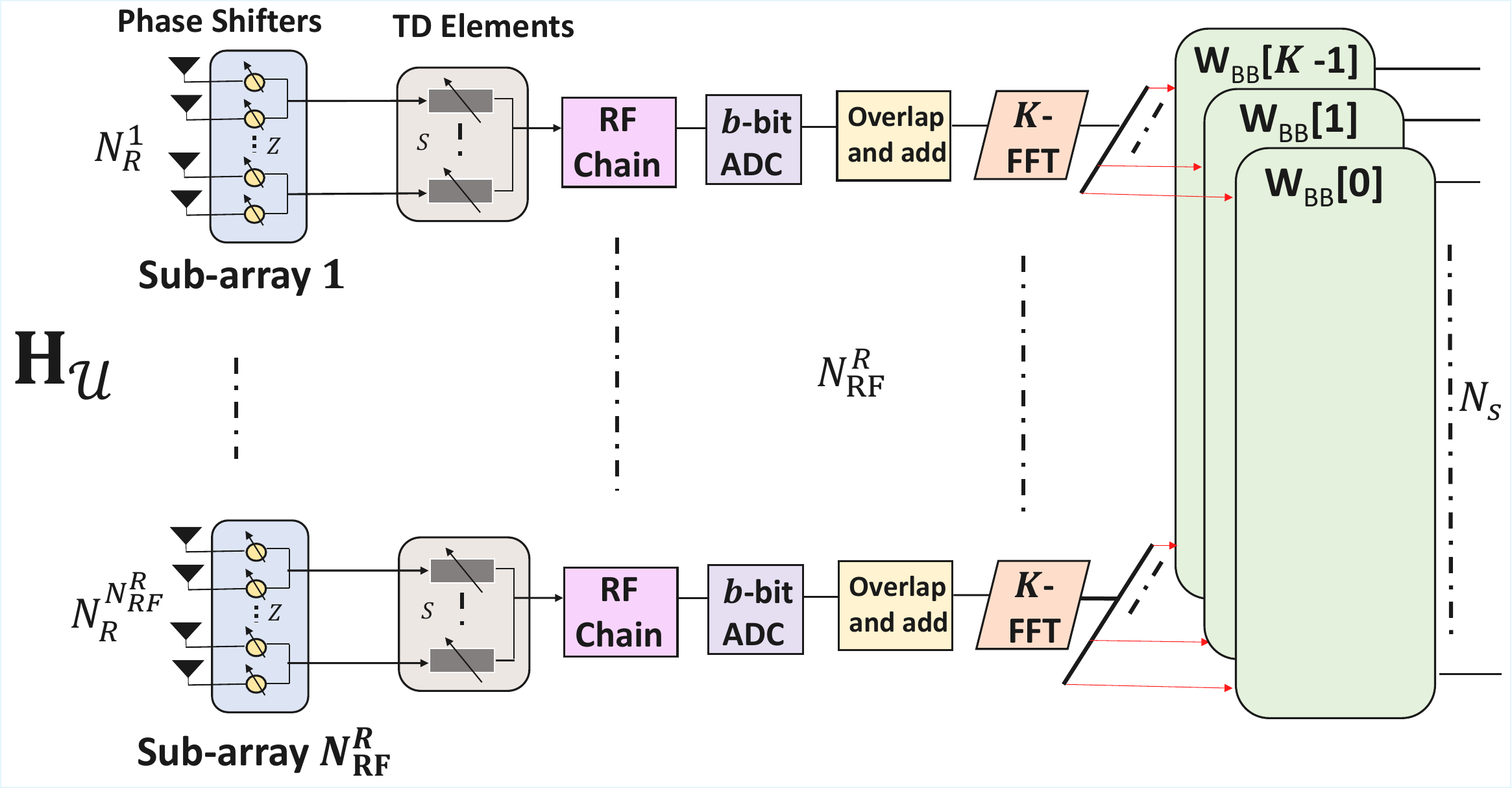}}
\vspace{-2mm}
\caption{Subarray-based combining with ADCs and TD elements}\vspace{-1 \baselineskip}
\label{TT_D}
\end{figure}
\vspace{-2mm}
\subsection{Generation of frequency-dependent beamformers}
\vspace{-1mm}
Let each RF chain is connected to $S$ time delay (TD) elements which are further connected with $Z = \frac{N_R^l}{S}$ PSs as shown in Fig. \ref{TT_D}. In addition, the array response vector $\tilde{\mathbf{a}}_R^l$ at $l$-th subarray given in Eq. \eqref{arrayresponsevector} be divided into $S$ different vectors given as
\begin{align}
    \tilde{\mathbf{a}}_{1,k} = \tilde{\mathbf{a}}_1 e^{-j 2 \pi f_k t_1^l}, \cdots ,\tilde{\mathbf{a}}_{S,k} = \tilde{\mathbf{a}}_S e^{-j 2 \pi f_k t_S^l}, \label{ar_ttd__}
\end{align}
where $\tilde{\mathbf{a}}_{(.)}^T \in \mathbb{C}^{Z \times 1}$ and $\tilde{\mathbf{a}}(\theta_k,f_k) = [\tilde{\mathbf{a}}_1^T, \cdots, \tilde{\mathbf{a}}_S^T]^T \in \mathbb{C}^{N_R^l \times 1}$. Let $\mathbf{t}^l = [t_1^l, t_2^l, \cdots, t_S^l]^T \in \mathbb{C}^{S \times 1}$ represent the delays corresponding to each RF chain. Therefore, the overall frequency-dependent phase shifter corresponding to the $k$-th subcarrier can be given as
\vspace{-2mm}
\begin{align}
    \tilde{\mathbf{a}}_k^l = \mathrm{blkdiag}\big(\tilde{\mathbf{a}}_1^T, \cdots, \tilde{\mathbf{a}}_S^T\big)e^{-j2 \pi f_k\mathbf{t}^l}. \label{arr_res}
\end{align}
Let $\mathbf{p}_k^l = e^{-j2 \pi f_k \mathbf{t}^l}$ represent the frequency-dependent PS. To maintain the directivity, we generate the frequency-dependent beam to share the same form as the array response vector $\tilde{\mathbf{a}}_{k}^l$, expressed as
\vspace{-2mm}
\begin{align}
    \mathbf{p}_k^l = [1, e^{-j \pi \upsilon_{k}^l }, \cdots, e^{-j \pi (S-1)\upsilon_{k}^l}]^T, \label{beam_vector}
\end{align}
where $\upsilon_{k}^l \in [-1,1]$ represents the directional rotation vector. Therefore, by adjusting $\upsilon_{k}^l$, the beams generated by $\tilde{\mathbf{a}}_{k}^l$ can be aligned with the target physical direction $\theta^l$ corresponding to all the subcarriers $K$. Moreover, the maximum array gain is achieved at an angle $\omega_{\mathrm{opt}}$, when
\begin{align}
    \omega_{\mathrm{opt}} = \underset{\theta}{\mathrm{arg \, max}} |(\tilde{\mathbf{a}}^l(\theta,f_c))^H \tilde{\mathbf{a}}^l(\theta_k,f_k)|,
\end{align}
which further reduces to 
\vspace{-2mm}
\begin{align}
    \omega_{\mathrm{opt}} = \frac{\theta^l}{\vartheta_k} - \frac{\upsilon_k^l}{Z\vartheta_k}, \label{optimal_angle}
\end{align}
whose derivation is provided in Appendix-\ref{appendix-c}. Thus, aligning the optimal beam steering angle with the actual physical direction yields $\upsilon_k^l = (1-\vartheta_k)Z\theta^l$, implying that the beam squint effect can be fully compensated using the TD elements. Additionally, it can be observed from Eq. \eqref{beam_vector}, the phase differences between adjacent TTD elements are equal and therefore the TD vector $\mathbf{t}^l$ should satisfy the following form which includes
\begin{align}
    \mathbf{t}^l = [0, nT_c, \cdots, (S-1)nT_c]^T, \label{time_delay}
\end{align}
where $T_c = \frac{1}{f_c}$ represents the time-period of the carrier frequency and $n$ represents number of periods corresponding to each RF chain. Substituting the TD value in the frequency-dependent PS expression, one obtains
\begin{align}
    \mathbf{p}_k^l = [1, e^{-j 2 \pi f_k n T_c}, \cdots, e^{-j 2 \pi f_k (S-1)n T_c}]^T,
\end{align}
and comparing with Eq. \eqref{beam_vector}, the time period obtained is $n = \frac{Z\theta^l}{2 \vartheta_k}(\vartheta_k-1)$. In addition, one can observe that the number of periods depends not only on the fixed phase shifters $Z$ and the physical direction $\theta^l$, but also on the relative frequency $\vartheta_k$. Substituting the value of $n$ in the TD vector, one obtains
\begin{align}
    t_s^l = (s-1)nT_c = (s-1)\frac{Z\theta^l}{2}T_c - (s-1)\frac{Z\theta^l}{2 \vartheta_k}T_c,
\end{align}\label{td_ly}
which represents the first term to be fixed $\tilde{t}_s^l = (s-1)\frac{Z\theta^l}{2}T_c, \, \forall \, 1<s<S,$ for all subcarriers while the second term can be realized by adding an extra phase shift. Revisiting Eq. \eqref{arr_res} and substituting the value of the TD vector, the overall array response vector can be determined as
\begin{align}
    \tilde{\mathbf{a}}_k^l &= \mathrm{blkdiag}(\tilde{\mathbf{a}}_1^T, \tilde{\mathbf{a}}_2^T, \cdots, \tilde{\mathbf{a}}_S^T)e^{-j 2 \pi f_k \mathbf{t}^l}, \notag \\ &= \mathrm{blkdiag}\big(\tilde{\mathbf{a}}_1^T, \tilde{\mathbf{a}}_2^T e^{j \pi Z \theta^l}, \cdots, \tilde{\mathbf{a}}_S^T e^{j \pi (S-1)Z \theta^l}\big)e^{-j2\pi f_k\mathbf{t}^l}, \notag
\end{align}
where the actual time delay $\tilde{\mathbf{t}}^l = [\tilde{t}_1^l, \tilde{t}_2^l, \cdots, \tilde{t}_S^l]$. Therefore, the generalized time delay corresponding to each RF chain can be derived as
\begin{align}
    \tilde{t}_s^l = \scalebox{0.99}{$\begin{cases}
        (s-1)\frac{Z\theta^l}{2}T_c, \quad\quad\quad\quad\quad\quad\quad\quad\:\:\: \theta^l \geq 0,  \\
        (S-1)|\frac{Z \theta^l}{2}|T_c + (s-1)\frac{Z\theta^l}{2}T_c, \quad\:\: \theta^l < 0. \label{final_TD}
    \end{cases}$}
\end{align}
Algorithm-\ref{precoder} outlines the design procedure for the frequency-dependent precoder. A similar procedure can be applied to obtain the hybrid combiner, where the optimal combining angles $\tilde{\Theta}_R$ are used as input. The design methodology is analogous to that presented in Algorithm-\ref{precoder}, with the only modification being the equivalent channel construction given by $\tilde{\mathbf{H}}_{\mathrm{eq}}[k] = \mathbf{W}_{\mathrm{RF}}^H[k]\mathbf{H}_{\mathcal{U}}[k] \tilde{\mathbf{F}}_{\mathrm{RF}}[k]\tilde{\mathbf{F}}_{\mathrm{BB}}[k]$ where $\tilde{\mathbf{F}}_{\mathrm{RF}}[k], \tilde{\mathbf{F}}_{\mathrm{BB}}[k]$ are constructed following the same procedure described in Section-\ref{MU_model}. Moreover, the RF combiner $\mathbf{W}_{\mathrm{RF}}[k]$ is formed by block-diagonally concatenating the array response vectors, as detailed in Section-\ref{MU_model}. Finally, the baseband combiner $\mathbf{W}_{\mathrm{BB}}[k]$ is obtained using the MMSE as $\mathbf{W}_{\mathrm{BB}}[k] = \tilde{\mathbf{H}}_{\mathrm{eq}}[k]\big(\tilde{\mathbf{H}}_{\mathrm{eq}}^H[k]\tilde{\mathbf{H}}_{\mathrm{eq}}[k]+\frac{UN_s}{\mathrm{SNR}}\mathbf{I}_{UN_s} \big)^{-1}$. The next section will discuss the simulation results.
\begin{algorithm}[t]
\DontPrintSemicolon 
\KwIn{Estimated channel $\{\widehat{\mathbf{H}}_u[k]\}_{k=1}^K,$ frequency independent precoder $ \mathbf{F}_{\mathrm{RF},u},$ dominant angle of departure $ \{\tilde{\Phi}_T[k]\}_{k=1}^K$}
 
\For{$k = 1: K$}{
\For{$\mathsf{i} = 1, 2, \cdots, N_{\mathrm{RF},u}^T$}
{
 $\tilde{\mathbf{a}}^{\mathsf{i}} = \mathbf{F}_{\mathrm{RF},u}[:,\mathsf{i}]; \: \theta^{\mathsf{i}} = \tilde{\Phi}_{T}(u,:)[\mathsf{i}]$

  \For{$s = 1,2,\ldots,S$}
 {
$\tilde{\mathbf{a}}^{\mathsf{i}}_s = \tilde{\mathbf{a}}^{\mathsf{i}}[1+(s-1)P:sP] e^{j \pi (S-1) P \theta^{\mathsf{i}}}$
 
Calculate the time delay $\tilde{t}_s^{\mathsf{i}}$ using equation \eqref{final_TD}

$\tilde{\mathbf{a}}_{s,k}^{\mathsf{i}} = \tilde{\mathbf{a}}_s^{\mathsf{i}} e^{-j 2 \pi f_k \tilde{t}_s^{\mathsf{i}}}$
}

$\mathbf{F}_{\mathrm{RF},u}[k](:,\mathsf{i}) = \big[(\tilde{\mathbf{a}}_{1,k}^{\mathsf{i}})^T, (\tilde{\mathbf{a}}_{2,k}^{\mathsf{i}})^T, \cdots,(\tilde{\mathbf{a}}_{S,k}^{\mathsf{i}})^T \big]^T$
 
}

$\tilde{\mathbf{H}}_{\mathrm{eq}}[k] = \mathbf{H}_u[k] \mathbf{F}_{\mathrm{RF},u}[k]$

$\left[\mathbf{U}_{\mathrm{eq}}[k] \mathbf{\Sigma}_{\mathrm{eq}}[k] \mathbf{V}_{\mathrm{eq}}^H[k]\right] = \mathrm{SVD}(\tilde{\mathbf{H}}_{\mathrm{eq}}[k])$

$\mathbf{F}_{\mathrm{BB},u}[k] = \mathbf{V}_{\mathrm{eq},[:,1:N_s^u]}[k]$
}

\textbf{return:~~}{$\mathbf{F}_{\mathrm{RF},u}[k], \mathbf{F}_{\mathrm{BB},u}[k]$}
\caption{TTD based frequency dependent precoder design}
\label{precoder}
\end{algorithm}
\vspace{-6mm}
{ \subsection{Effect of Insertion Loss}\label{ins_l_ss}
\vspace{-1mm}
The physical implementations of TTD elements such as switched transmission lines or optical delay lines introduce insertion loss (IL) that generally increases with delay length. Therefore the loss should be modeled as an amplitude attenuation factor within the hybrid beamforming matrix. Moreover modifying \eqref{ar_ttd__}, the modified array response vector with insertion loss can be rewritten as $\mathbf{\tilde{a}}_{s,k}=\mathbf{\tilde{a}}_s L_{s}e^{-1j2\pi f_kt_{s}^l}$, with $L_s$ is attenuation factor for $s$-th TTD at $l$-th RF chain which is given as $L_s=10^{\left(\frac{-L_s(dB)}{20}\right)}$. Note that, we have considered uniform loss for all TTD elements $L_1=L_2=\dots=L_S$ which is in line with the existing literature \cite{yan2022energy}. Therefore Eq. \eqref{arr_res} with insertion loss, can be re-written as $\mathbf{\tilde{a}}_k^l=\mathrm{blkdiag}\left(\mathbf{\tilde{a}}_1^T,\mathbf{\tilde{a}}_2^T,\dots,\mathbf{\tilde{a}}_S^T\right)Le^{-j2\pi f_k\mathbf{t}^l}$. Finally, the modified array response vector corresponding to each TTD delay is given as
\begin{align}
\mathbf{\tilde{a}}_k&=\mathrm{blkdiag}\left(\mathbf{\tilde{a}}_1^T,\mathbf{\tilde{a}}_2^T,\dots,\mathbf{\tilde{a}}_S^T\right)Le^{-j2\pi f_k \mathbf{t}},\\
&=\mathrm{blkdiag}\left(\mathbf{\tilde{a}}_1^T,\mathbf{\tilde{a}}_2^Te^{j\pi Z\theta},\dots,\mathbf{\tilde{a}}_S^Te^{j\pi(S-1)Z\theta}\right)Le^{-j2\pi f_k \mathbf{t}}. \notag
\end{align}
}
\vspace{-10mm}
{ \subsection{Effect of quantized delay}
The continuous delay values $t$ corresponding to each $s$-th subarray as described in Eq. \eqref{final_TD}
are quantized to $\widehat{\tilde{t}_s} = q_b(\tilde{t}_s)$, where $q_b(\cdot)$ rounds to the nearest discrete level among $2^{\tilde{d}}$ possible values. The quantization step size is $\Delta = \frac{t_{\max}}{2^{\tilde{d}}-1}$, resulting in a maximum quantization error bounded by $\mathrm{er} = |\tilde{t}_s - \widehat{\tilde{t}}_s| \leq \Delta/2$. Here, $\tilde{d}$ denotes the total number of finite bits used for quantization, and each delay value $\tilde{t}_s$ lies within the interval $[0, \frac{NT_c}{2}]$, implying that the maximum delay is $t_{\max} = \frac{NT_c}{2}$. This quantization error induces phase deviation $\Delta\phi=2\pi f_k \mathrm{er}$ that vary across frequencies, further contributing to beam squint in wideband systems.}
\vspace{-3mm}
\section{Simulation Results} 
\label{simulation}
This section provides a performance analysis for the proposed variational Bayesian inference and the TTD-based transceiver design. Table-\ref{simulation_para} outlines the key numerical parameters used in the simulation study. Additionally, the inter-antenna spacing between the AEs is set to be $\frac{\lambda}{2}$. The material parameters of various scatterers, as listed in Table-\ref{materials_para}, are used in the simulation. Note that in the THz band, the refractive index exhibits frequency-dependent behavior, which causes these materials to interact differently with electromagnetic waves across the frequency spectrum \cite{piesiewicz2007properties}. The Signal-to-Noise Ratio (SNR) in decibels (dB) is calculated as $\mathrm{SNR} = 10 \mathrm{log}_{10}(\frac{1}{\sigma_n^2})$. The transmitting and receiving grid-sizes are considered to be $G_{T,u} \geq 2N_{T,u}$ and $G_R \geq 2N_R$ \cite{gonzalez2018channel} for on-grid scenario. { Note that, although the derived condition for $G_R^l$ in Section-\ref{Channel_mod}(B) provides a sufficient bound for the validity of the first-order Taylor approximation, strictly satisfying this condition would require a significantly large grid size, leading to increased computational complexity. Therefore, a moderate grid resolution is adopted, and the Taylor expansion effectively compensates for the residual off-grid mismatch, achieving a favorable trade-off between estimation accuracy and computational complexity. For $G_R^l = 4 N_R^l$, the resulting angular mismatch is approximately $\approx 0.1$ rad, which is sufficiently small to achieve a favorable trade-off between estimation accuracy and computational complexity \cite{yang2012off}. A similar assumption is adopted at the transmitter side.} The frequency independent phase shifters are modeled similar to \cite{garg2024angularly} given as
\vspace{-2mm}
\begin{align}
    \mathbf{F}_{\mathrm{RF},m,u}(a,b) = \frac{1}{\sqrt{N_{T,u}}} e^{j\varkappa_{a,b}}, \mathbf{W}_{\mathrm{RF},m} = \frac{1}{\sqrt{N_R}}e^{j\psi_{a,b}}, \notag
\end{align}
where the phases $\varkappa_{a,b}, \psi_{a,b}$ are randomly sampled with uniform probability as $\mathcal{F} = \Big\{0,\frac{2\pi}{2^{N_Q}},\cdots,\frac{(2^{N_Q}-1)}{2^{N_Q}} \Big\}.$ { Note that the considered RRC pulse shaping parameters as described in Table-\ref{simulation_para} are consistent with standard wideband system configurations \cite{venugopal2017channel} and align with practical implementations such as IEEE 802.11ad \cite{kumari2017ieee}. Moreover, the free-space molecular absorption loss is computed using the standard HITRAN database \cite{hill}. The database comprises both experimentally measured and theoretically derived spectroscopic parameters \cite{hill}. Combined with radiative transfer theory, these parameters are utilized to accurately characterize molecular absorption losses in THz band as detailed in \cite{jornet2011channel}.}
\vspace{-3mm}
{\subsection{GMM based MU AoA/ AoD generation}
We harness a GMM for generating the AoAs/AoDs, as it provides an accurate representation of the angular characteristics in the THz band \cite{lin2015adaptive}. To generate angles corresponding to distinct users, the mean angle for each user is drawn from a uniform distribution, $\tilde{\theta}_u \sim \mathcal{U}[-180^\circ, 180^\circ]$, where $\tilde{\theta}_u$ denotes the mean angle of the $u$-th user.
\begin{table}
\vspace{-8pt}
\centering
{ \caption{Simulation parameters for the hybrid transceiver design and sparse estimation \vspace{-1\baselineskip}}}
\label{simulation_para}
\resizebox{0.45\textwidth}{!}{%
\begin{tabular}{|l|c|}\hline
\textbf{Parameter} & \textbf{Value} \\ \hline \hline
$\#$ of TAs/ RAs at UE ($N_{T,u}$, $N_R$)  & $4, 64$ \\\hline
$\#$ of TAs/ RAs RF chains ($N_{\mathrm{RF}}^T$, $N_{\mathrm{RF}}^R$)  & $2, 8$ \\ \hline
$\#$ of time delay per RF chain ($S$)  & $2$ \\ \hline
$\#$ of antennas at each subarray ($N_R^l$) &  $8$ \\ \hline
$\#$ of subcarriers ($K$) &  $128$ \\ \hline
$\#$ of pilot blocks ($M$) & $20$ \\ \hline
$\#$ of data vectors ($N_d$)  & $100$ \\ \hline
$\#$ of Users ($U$)  & $3$ \\ \hline
$\#$ of NLoS components ($N_{\text{NLoS}}$) &  $3$ \\ \hline
$\#$ of delay taps ($D$) &  $3$ \\ \hline
Molecular Absorption loss ($K_{\mathrm{abs}}$) &  $0.015$ $\mathrm{m}^{-1}$ \\ \hline
Transmit Antenna Gain at each UE ($\mathfrak{D}_{T,u}$) &  $8 \: \text{dBi}$ \\ \hline
Receive Antenna Gain at each subarray ($\mathfrak{D}_R^l$) &  $12 \: \text{dBi}$ \\ \hline
Transmit angular grid size at UE ($G_{T,u}$) & $8$ \\ \hline
{ Receive angular grid size at each subarray ($G_R^l$)} & { $32$} \\ \hline
Operating frequency ($f_c$), Bandwidth ($B$) & $0.65 \: \text{THz}$, $5 \: \text{GHz}$ \\\hline
Transmission distance ($d$) & $15 \text{m}$ \\\hline
Angle quantization parameter ($N_Q$) & $4$ \\\hline
ADC Resolution ($\tilde{\xi}$), Constellation & $4$-\text{bit}, $8$-\text{PSK} \\\hline
{ Roll-off factor for RRC-PSF} & {$0.80$} \\\hline
{ Upsampling factor} & { $16$} \\ \hline
Bayesian Inference Threshold ($\varepsilon, N_\mathrm{max}$) & $2,30$ \\ \hline
\hline
\end{tabular}
}
\vspace{-1.2\baselineskip}
\end{table}
\begin{table}
\centering
{\caption{List of materials used for simulation environment \cite{piesiewicz2007scattering},\cite{piesiewicz2007properties}\vspace{-1\baselineskip}}} \label{materials_para}
\resizebox{0.42\textwidth}{!}{%
\begin{tabular}{|l|c|c|c|}\hline
\textbf{Material Type} & $\sigma_r (\mathrm{in \: mm})$ & $\chi (\mathrm{in \: cm^{-1}})$ & $\eta$ \\\hline
Polycarbonate (PC) & $0$ & $23$ & $1.52$  \\\hline
Polystyrene (PS) & $0.002$ & $2$& $1.6$ \\\hline
Polyvinyl chloride (PVC) & $0.028$ & $19$ & $1.68$  \\\hline
Plaster s1 & $0.05$ & $10$ & $2$ \\\hline
Gypsum plaster & $0.13$ & $38$ & $1.4$  \\\hline
Plaster s2 & $0.15$ & $10$ & $2$  \\\hline
\end{tabular}%
}
\vspace{-0.8\baselineskip}
\end{table}
\begin{figure*}[t]
	\centering
   \subfloat[]{\includegraphics[scale=0.3]{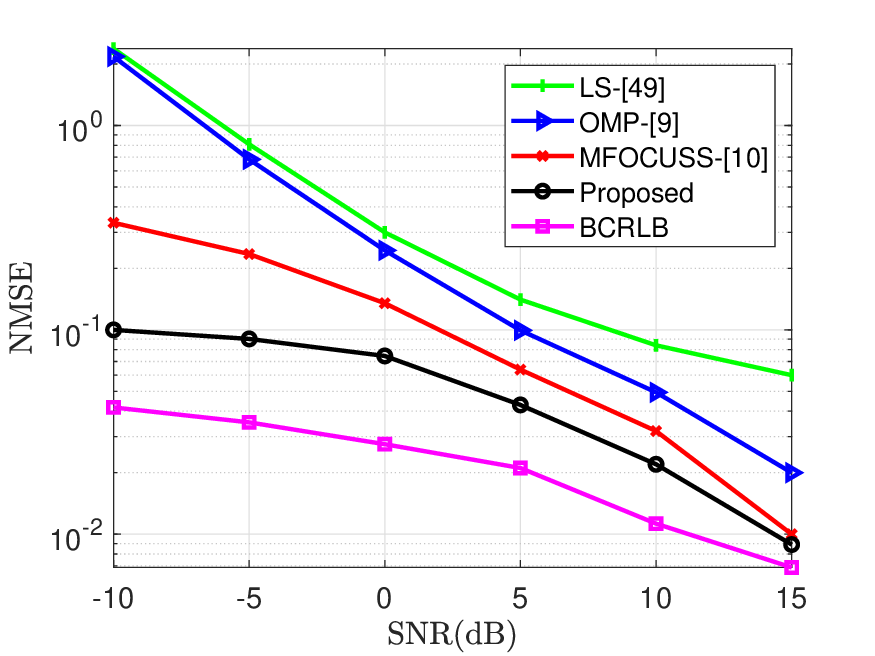}}
	\hfil
\hspace{-10pt}\subfloat[]{\includegraphics[scale=0.3]{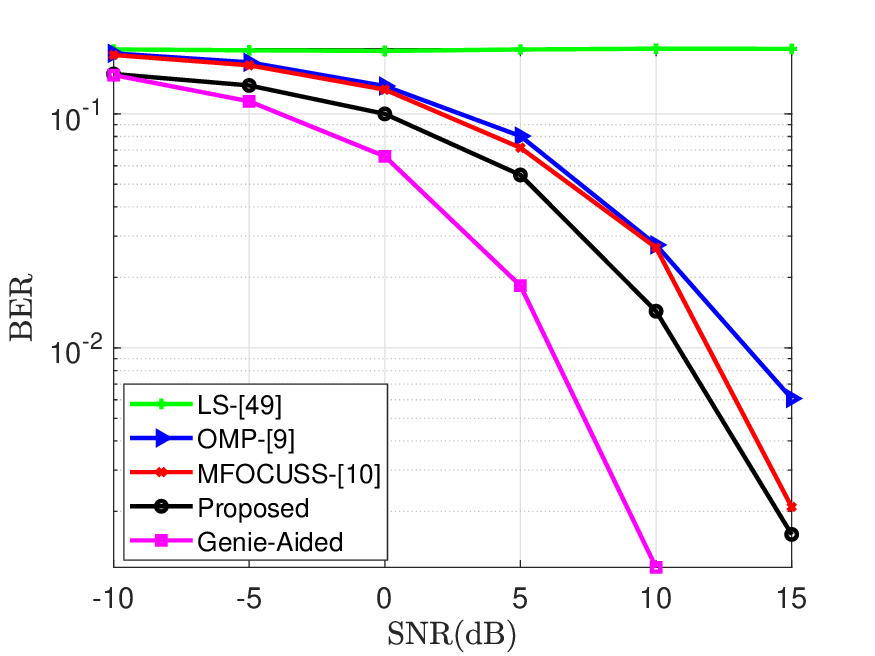}}
	\hfil
\hspace{-10pt}\subfloat[]{\includegraphics[scale=0.3]{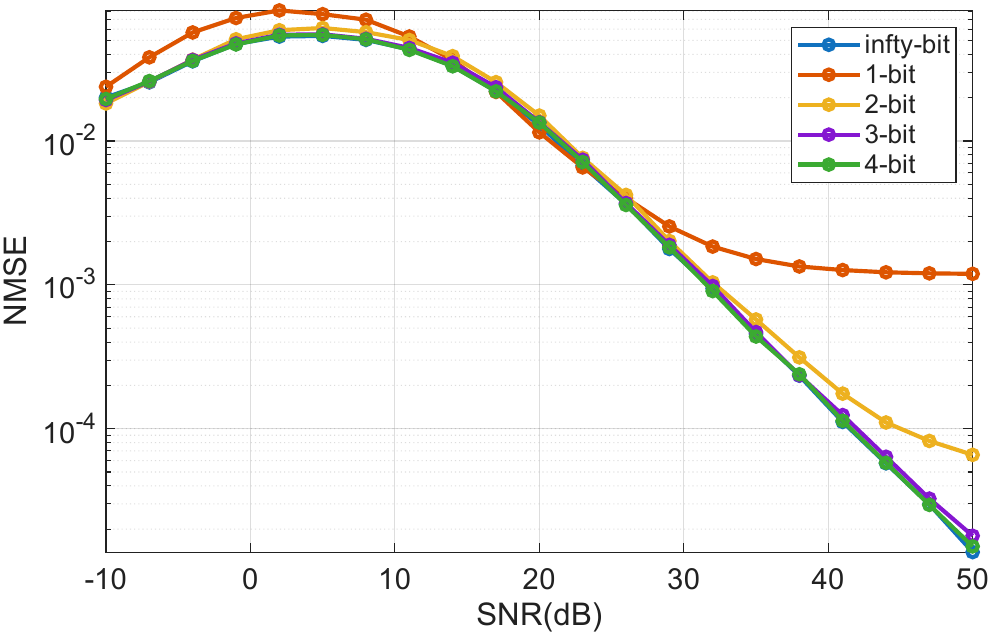}}
\hfil
\hspace{-10pt}\subfloat[]{\includegraphics[scale=0.3]{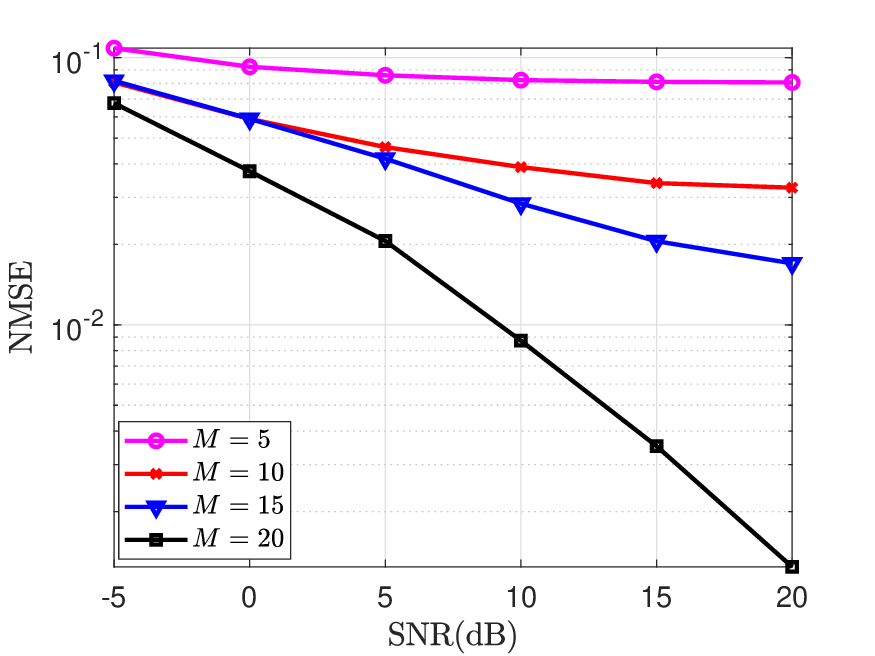}}
	\caption{$ \left(a\right) $ NMSE vs SNR comparison for the proposed and existing state-of-the-art approaches $ \left(b\right) $ BER vs SNR comparison for the proposed and existing state-of-the-art approaches { $ \left(c\right)$ Effect of low-resolution ADCs on proposed Bayesian inference $ \left(d\right) $ NMSE vs SNR comparison by varying pilot blocks $M$.}\vspace{-7.2\baselineskip}}\label{NMSE}
\end{figure*}
Around each mean, the angular spread is modeled by a Gaussian kernel of variance $\check{\sigma}^2$, where $\check{\sigma}^2$ is chosen according to the measurements reported in \cite{priebe2011aoa} (Table III). Thus, the AoA/AoD distribution using a two-component GMM can be expressed as
\vspace{-3mm}
\begin{align}
    \text{GMM}(x_u) = \sum_{\iota=1}^2 a_{u,\iota} \mathcal{N}(x_u|\tilde{\theta}_{u,\iota}, \check{\sigma}^2),
\end{align}
where $a_{u,\iota}$ represent the mixture weight. They are randomly drawn from a uniform distribution $\{a_{u,1},a_{u,2}\} \sim \mathcal{U}(0,1)$ satisfying $a_{u,1}+a_{u,2} = 1,$ ensuring a valid convex combination. Furthermore, to avoid angular overlap among different users, we impose a \textit{minimum separation constraint of $|\tilde{\theta}_u - \tilde{\theta}_v| \geq d_{\text{min}} \, \forall \, u \neq v.$}
\begin{figure*}
	\centering
   \subfloat[]{\includegraphics[scale=0.3]{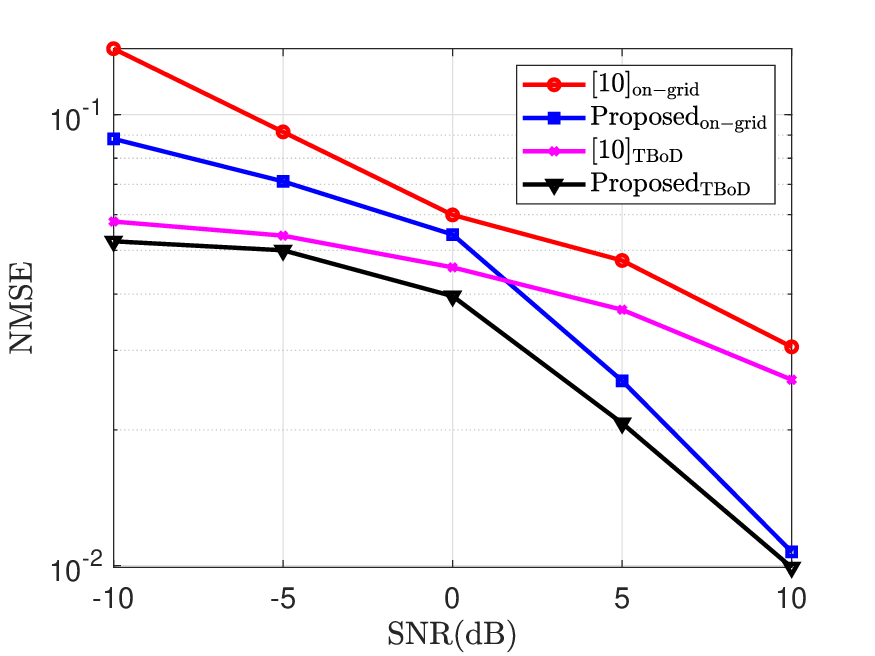}}
	\hfil
\hspace{-10pt}\subfloat[]{\includegraphics[scale=0.3]{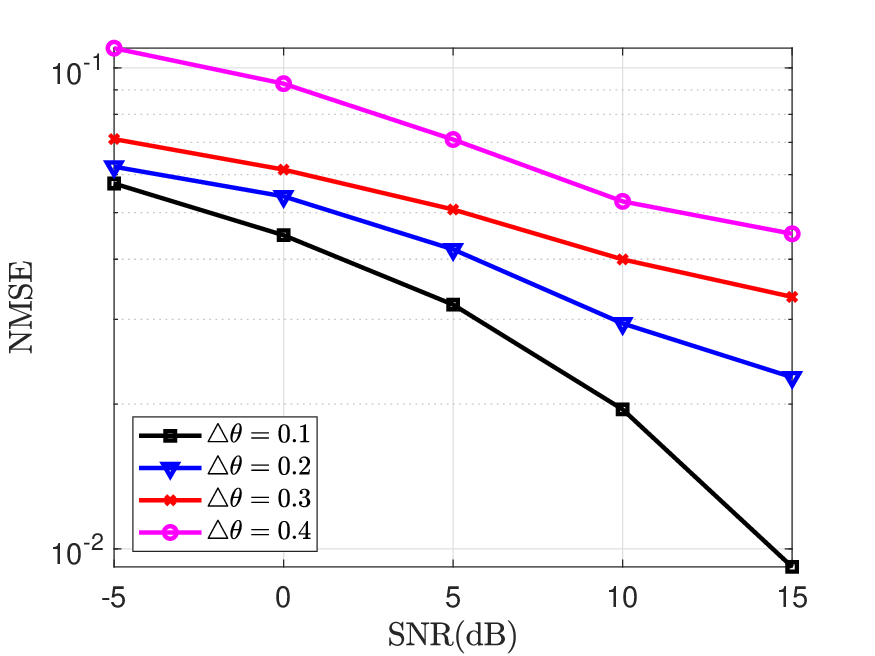}}
	\hfil
\hspace{-10pt}\subfloat[]{\includegraphics[scale=0.3]{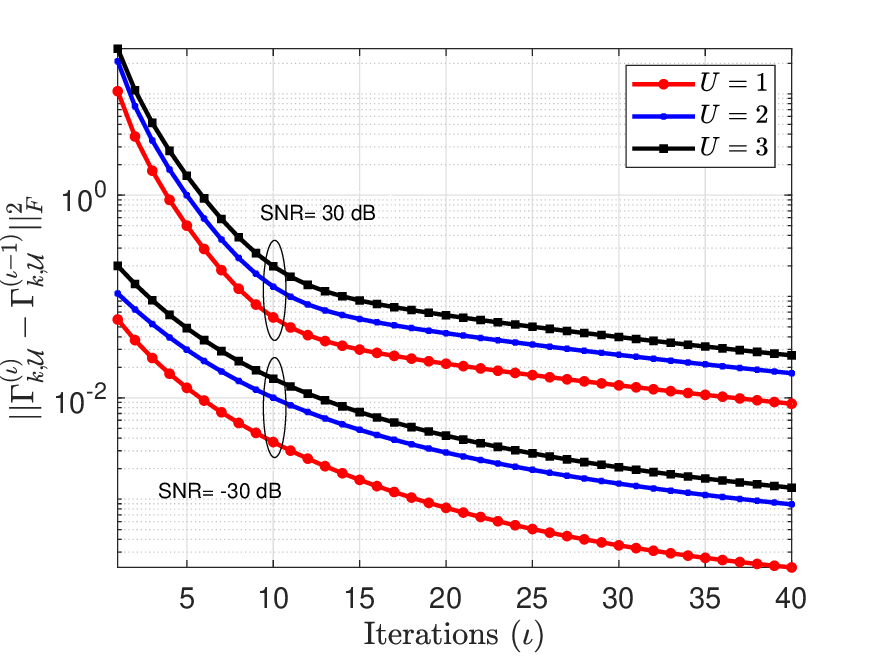}}
	\hfil
\hspace{-10pt}\subfloat[]{\includegraphics[scale=0.3]{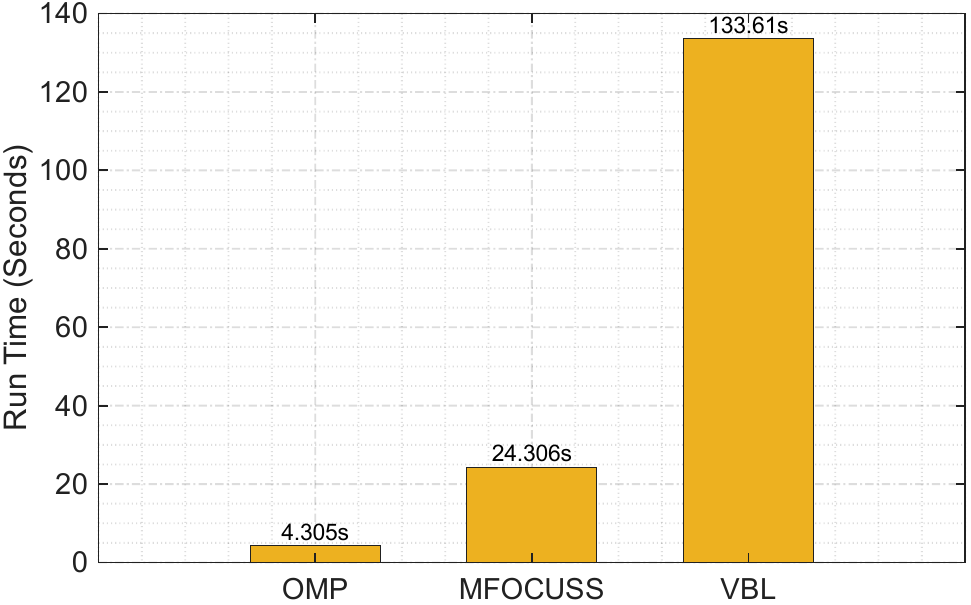}}
	\caption{$ \left(a\right) $ NMSE vs SNR comparison for the uniform and Taylor based off-grid dictionary $ \left(b\right) $ NMSE vs SNR comparison by varying the offset angles $\triangle \theta$ $ \left(c\right) $ $\parallel \boldsymbol{\Gamma}_{k,\mathcal{U}}^{(\iota)} - \boldsymbol{\Gamma}_{k,\mathcal{U}}^{(\iota-1)}\parallel^2_\mathcal{F}$ vs number of iterations by varying the users at $\mathrm{SNR}=[30,-30]\mathrm{dB}$ { $\left(d\right)$  Run time comparison for the considered sparse frameworks.}\vspace{-2.9\baselineskip} }\label{off-grid}
\end{figure*}
To generate angles corresponding to distinct users, the mean angles $x_u$ for AoA/AoD are selected such that all angles lie within $\tilde{\theta} \in \mathcal{U}[-180^\circ, 180^\circ]$ with a minimum angular separation of $d_{\mathrm{min}} = 5^\circ$. The angular spread around each mean angle are uniform across all users, with parameters $\check{\sigma}_1^2 = 108^\circ$ and $\check{\sigma}_2^2 = 314^\circ$. The NMSE is given by $\mathrm{NMSE} = \frac{\sum_{k=0}^{K-1}\parallel \widehat{\mathbf{H}}[k]-\mathbf{H}[k]\parallel_\mathcal{F}^2}{\sum_{k=0}^{K-1}\parallel \mathbf{H}[k] \parallel_\mathcal{F}^2}$. Note that, for all simulations, we employ the RRC-PSF-based dual-wideband channel model, and for comparative evaluation, we also consider the rectangular pulse shaping filter (Rect-PSF)-based model described in \cite{dovelos2021channel}. A trade-off analysis between the two channels are also presented in Section-VI(B).\\
{ \underline{\textit{Initialization of parameters for variational Bayesian inference}:} To ensure non-informative priors in the probabilistic model, the hyperparameters $\{\mathsf{m},\mathsf{c}\}_{t=1}^{G_RG_T}$ and $\{\mathsf{w},\mathsf{e}\}$ are all set to $10^{-6}$ \cite{tzikas2008variational}. The expectations $\mathbb{E}\{\boldsymbol{\Gamma}_{k,\mathcal{U}}\}= \mathbf{I}_{G_RG_T}, \, \mathbb{E}\{\flat\} = \frac{\mathsf{w}}{\mathsf{e}} = 1$ are initialized using their corresponding prior.} Note that, considering small prior values does not imply that the priors are purposeless or can be omitted. It means they have minimal influence on the posterior or variational distributions. Therefore, during inference, the information provided by the observations, encapsulated in the likelihood function, is combined with the priors via Bayes rule to form the posterior distribution, which then incorporates the observed data and becomes informative.
\vspace{-4mm}
\subsection{Analysis of Channel Estimation}
\vspace{-1mm}
This section presents the performance analysis of the proposed variational Bayesian inference method and compares it with the existing state-of-the-art techniques. Fig. \ref{NMSE}(a) illustrates the NMSE performance as a function of SNR. The poor performance of the least squares (LS) method \cite{haupt2010toeplitz} is attributed to its inability to incorporate sparsity priors into the estimation process, which poses a significant limitation compared to compressive sensing (CS)-based approaches. Moreover, the orthogonal matching pursuit (OMP) algorithm \cite{karabulut2004sparse} suffers from both structural and convergence errors. Specifically, using a low stopping threshold results in a large number of iterations and excessive non-zero components, introducing structural error. Conversely, a high stopping threshold leads to fewer iterations and fails to capture all dominant channel components, resulting in convergence error. Moreover, the MMV Focal Underdetermined System Solver (MFOCUSS) algorithm \cite{wipf2007empirical} often converges to suboptimal local minima, resulting in convergence errors. In contrast, the proposed variational Bayesian inference assumes parameterized prior distributions and optimizes the variational parameters to approximate the posterior distribution. It eliminates the need for manual tuning or regularization parameters and remains robust to the choice of the dictionary matrix. Additionally, the  Bayesian inference, provides statistically grounded posterior distributions over the unknown variables, enabling uncertainty quantification which are not supported by conventional point estimate based methods, thereby enabling the joint estimation of all unknown variables and making it well-suited for diverse propagation environments. Further, the performance of the proposed technique closely approaches the BCRLB, which is particularly noteworthy given that the BCRLB is derived under the idealized assumption of perfect AoA/AoD knowledge. Fig. \ref{NMSE}(b) depicts the accuracy of data detection achieved by employing the Minimum Mean Squared Error (MMSE) receiver with the proposed Bayesian inference, along with other conventional sparse estimation based techniques and the Genie-aided detector as benchmarks. As illustrated, the resulting BER decreases consistently with increasing SNR, which closely aligns with the NMSE results presented previously, underscoring the high quality of the CSI estimated by the Bayesian inference. Additionally, the BER of the proposed channel learning approach closely matches the hypothetical Genie-aided detector at higher SNRs, highlighting its efficacy and robustness in realistic scenarios.

{ Fig. \ref{NMSE}(c) illustrates the performance across different ADC resolutions over a wide SNR range. Specifically, at low SNR values, all ADC resolutions exhibit similar performance, indicating that thermal noise is the dominant impairment. In this regime, the impact of quantization is negligible, and even $1$-bit ADCs achieve performance comparable to higher-resolution counterparts. However, as the SNR increases, the effect of thermal noise diminishes, and quantization distortion becomes the limiting factor \cite{yang2021secure}. This is evident from the performance saturation observed for low-resolution ADCs, particularly in the $1$-bit case, where an error flooring appears at high SNR which is inline with the existing works \cite{wang2017bayesian}. In contrast, higher-resolution ADCs ($3$-bit and $4$-bit) continue to improve with increasing SNR and closely approach the $\infty$-resolution benchmark, indicating reduced quantization distortion. The $2$-bit ADC achieves a favorable trade-off, exhibiting near-optimal performance over a wide SNR range while maintaining low hardware complexity \cite{xu2018performance}. These observations validate the effectiveness of the Bussgang-based modeling in accurately capturing quantization effects across different operating regimes. It clearly demonstrate the transition between thermal and quantization noise-limited regimes. Fig. \ref{NMSE}(d) presents the NMSE versus SNR performance for different numbers of pilot blocks $M$. As $M$ increases, the NMSE consistently improves due to the availability of more measurements, which strengthens the Bayesian inference process by more accurately capturing the sparse channel structure. This results in reduced estimation uncertainty and improved robustness to noise. }
\begin{figure*}
	\centering
   \subfloat[]{\includegraphics[scale=0.295]{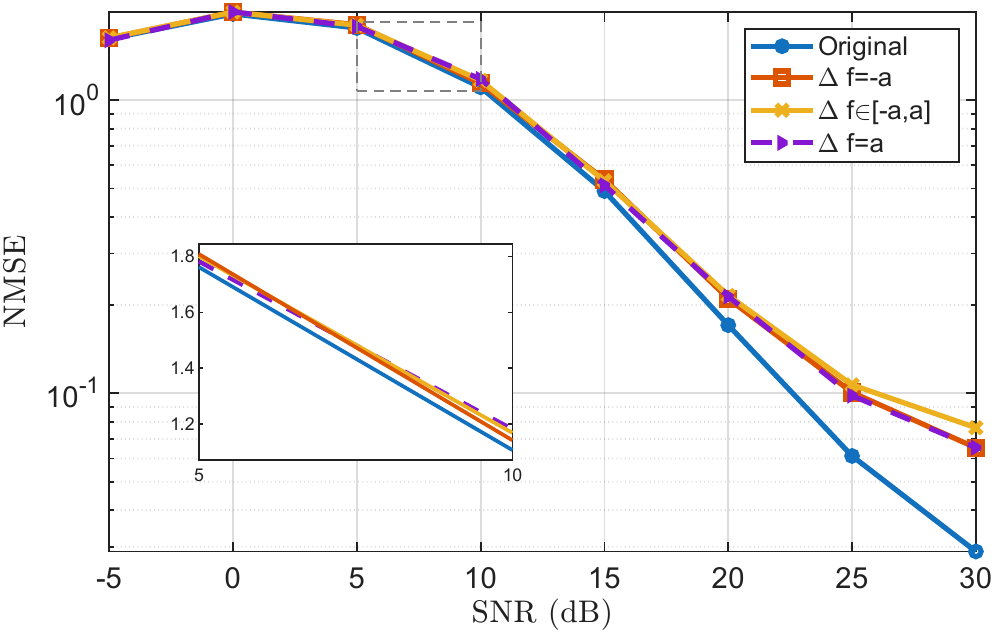}}
	\hfil
\hspace{-10pt}\subfloat[]{\includegraphics[scale=0.295]{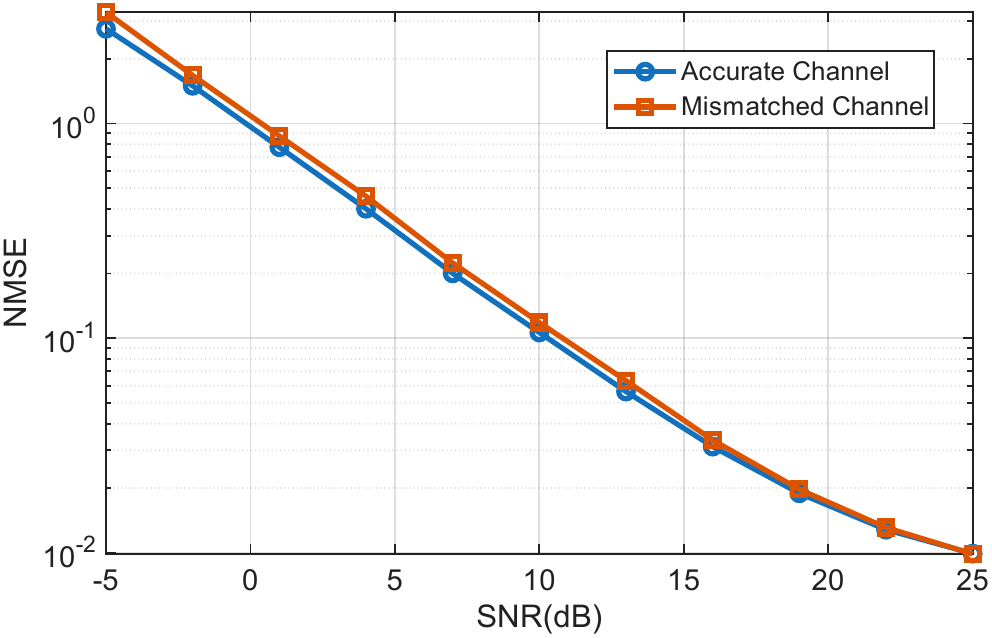}}
	\hfil
\subfloat[]{\includegraphics[scale=0.295]{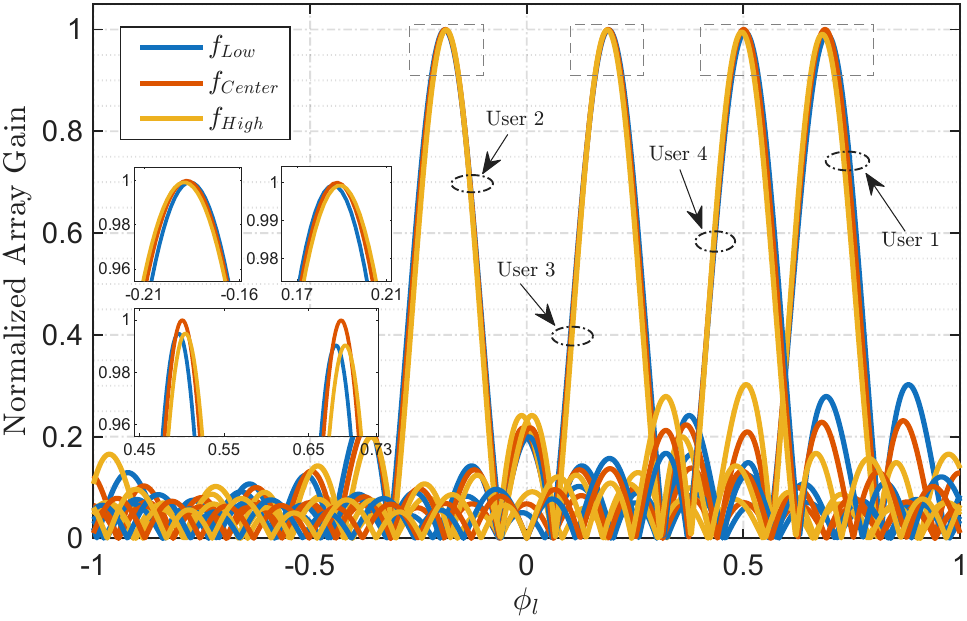}}
    	\caption{{ $ \left(a\right) $  NMSE vs SNR comparison by varying CFO values $ \left(b\right) $  NMSE vs SNR comparison for true and mismatched channel} { $ \left(c\right) $ Normalized array gain vs physical direction corresponding to all the users} }\label{new_simu}\vspace{-2\baselineskip} 
\end{figure*}
\begin{figure*}
	\centering
\hspace{-5pt}\subfloat[]{\includegraphics[scale=0.3]{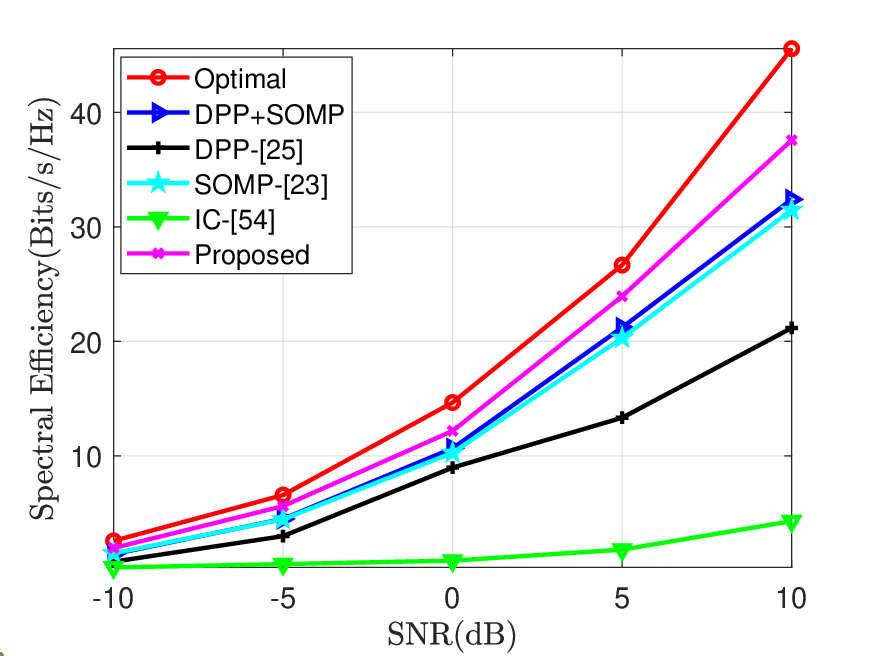}}
	\hfil
\hspace{-5pt}\subfloat[]{\includegraphics[scale=0.295]{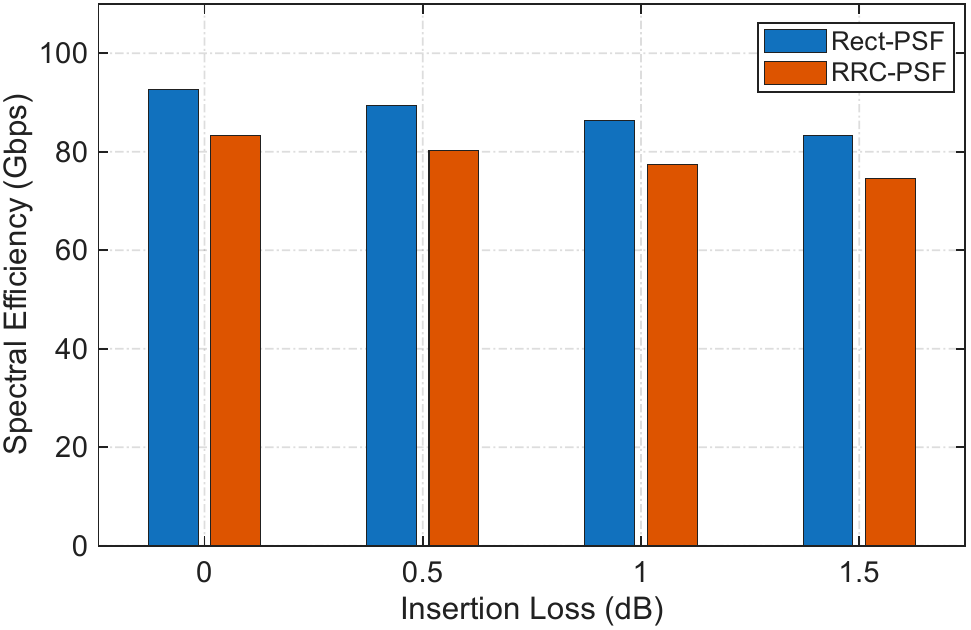}}
	\hfil
\hspace{-5pt}\subfloat[]{\includegraphics[scale=0.28]{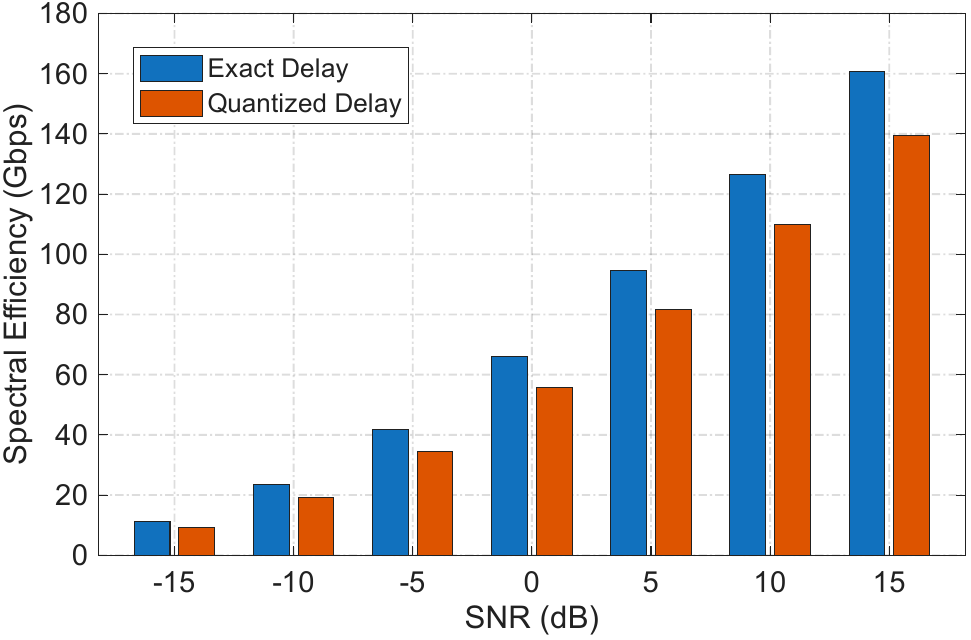}}
	\hfil
\hspace{-5pt}\subfloat[]{\includegraphics[scale=0.3]{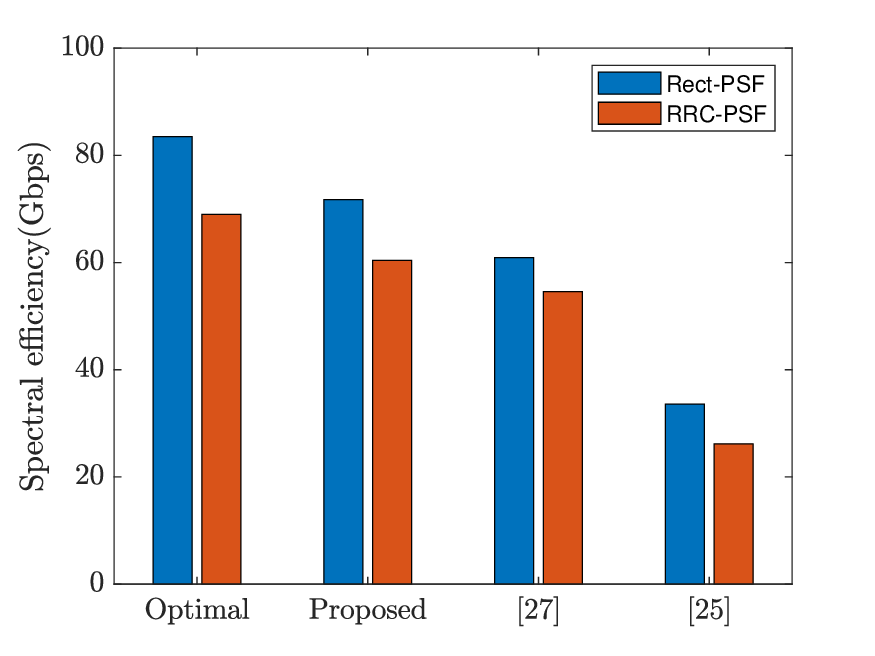}}
	\caption{ $ \left(a\right) $ Spectral efficiency vs SNR for the proposed and existing state-of-the-art approaches { $ \left(b\right)$ SE vs insertion loss for Rect-PSF and RRC-PSF based dual-wideband channels $\left(c\right)$ SE vs SNR for exact and quantized delay }$\left(d\right)$ Spectral efficiency comparison for the Rect-PSF and RRC-PSF based dual-wideband channels.\vspace{-3.3\baselineskip}}\label{SE}
\end{figure*}

Fig. \ref{off-grid}(a) evaluates the comparative performance of the proposed TBsD matrix against the conventional on-grid dictionary. As observed, the TBsD achieves superior results compared to the on-grid approach for both the M-FOCUSS \cite{wipf2007empirical} and Bayesian inference, primarily because it can effectively capture continuous angular variations. While the on-grid dictionary is inherently limited by discretized angular grids, leading to inevitable basis mismatch errors, the off-grid TBsD circumvents this limitation by providing a more refined and flexible angular representation with enhanced performance. Fig. \ref{off-grid}(b) compares the NMSE versus SNR performance for various values of the off-grid angular offset parameter $\triangle \theta$. As observed from the results, selecting a smaller value of $\triangle \theta$ allows the off-grid dictionary to represent continuous angular deviations more precisely, thus significantly enhancing channel estimation accuracy. Conversely, increasing $\triangle \theta$ leads to coarser angular representation, causing the dictionary to resemble an on-grid structure, which adversely affects estimation accuracy due to increased basis mismatch errors. Therefore, the proposed algorithm demonstrates robust performance. { Fig. \ref{off-grid}(c) empirically validates the convergence behavior, showing that the algorithm reaches the target NMSE of about $0.5$ in $\approx 5$ iterations for a single user and $\approx 8$ iterations for three simultaneous users at $\mathrm{SNR} = 30$ dB. The results demonstrate a monotonic decrease in the estimation error with iterations, indicating stable convergence of the algorithm. Moreover, the convergence behavior is consistent across different SNR values and system configurations, with faster convergence observed at higher SNR due to improved observation quality. Furthermore, the proposed algorithm consistently converges to solutions approaching the BCRLB as represented in Fig. \ref{NMSE}(a), indicating that the stationary point represents the global ELBO optimum as discussed in Section-\ref{convergence_VB}. Fig. \ref{off-grid}(d) compares the computational runtime of OMP, MFOCUSS and the proposed Bayesian inference. It is observed that OMP exhibits the lowest runtime of 4.305 s, followed by MFOCUSS with 24.306 s, while the proposed Bayesian inference incurs a higher computational cost of 133.61 s which is inline with the computational complexity. The increased runtime of Bayesian inference is attributed to its iterative procedure, which involves posterior updates and hyperparameter estimation at each iteration, leading to increased computational overhead. However, this additional complexity enables more accurate recovery of sparse channel components and improved robustness compared to conventional greedy and reweighting-based methods. Thus, there is a clear trade-off between the accuracy and runtime.}
\vspace{-2mm}
{ \subsection{Performance under imperfect prior information and model mismatch}
\vspace{-1mm}
In order to realize the system with imperfect prior information, we have considered the impact of carrier frequency offset (CFO) into the received signal. In this regard, the normalized CFO $\Delta f$ is defined as $\Delta f=\Delta f_{\mathrm{actual}}\cdot T_s$ where $T_s$ denotes the sampling period and $f_s$ is the sampling frequency. Moreover, the CFO is uniformly distributed between $\frac{1}{-2T_s}$ and $\frac{1}{2T_s}$ and considered as adopted in \cite{rodriguez2019channel}. Thus, for the $m$th training block, the phase rotation matrix at the $k$th subcarrier is defined as
\vspace{-2mm}
\begin{equation}
\boldsymbol{\complement}_{k,m}=\exp\left(j2\pi\Delta f(k-1)\frac{m}{K}\right)\cdot\mathbf{I}_{N_R},
\end{equation}
where $\exp(j 2 \pi \Delta f(k-1)\frac{m}{K})$ represents the scalar phase factor. This phase rotation matrix acts multiplicatively on the beam-squint effected channel $\mathbf{H}[k]$, yielding the CFO-corrupted effective channel as
\vspace{-2mm}
\begin{equation}
    \mathbf{H}_{\mathrm{eff}}[k]=\boldsymbol{\complement}_{k,m}\mathbf{H}[k].
\end{equation}
Therefore, Eq. \eqref{sys-mod} after incorporating the CFO, can be re-written as
\vspace{-2mm}
\begin{equation}
    \begin{aligned}
 \mathbf{y}_m[k]\approx &\underbrace{\left(\tilde{\boldsymbol{\mathfrak{P}}}_m^T[k]\tilde{\mathbf{F}}_{\mathrm{BB},m}^T\tilde{\mathbf{F}}_{\mathrm{RF},m}^T\right)\otimes\left(\mathbf{W}_{\mathrm{BB},m}^H[k]\boldsymbol{\Lambda}\mathbf{W}_{\mathrm{RF},m}^H\boldsymbol{\complement}_{k,m}\right)}_{\tilde{\boldsymbol{\Psi}}^{\mathrm{CFO}}_{\mathcal{U},m}[k] \in \mathbb{C}^{N_s \times N_TN_R}}  \\ &\quad\quad\quad\quad\quad \mathrm{vec}(\mathbf{H}_{\mathcal{U}}[k])+\boldsymbol{\eta}_m[k],
\end{aligned}
\end{equation}
where the modified sensing matrix $\tilde{\boldsymbol{\Psi}}^{\mathrm{CFO}}_{\mathcal{U},m}[k]$ considers the CFO phase rotation into the effective measurement operator. Therefore, after concatenation over $M$ blocks, the received signal with CFO is given as
\vspace{-2mm}
\begin{equation}
    \mathbf{y}[k]\approx\boldsymbol{\Psi}_{\mathcal{U}}^{\mathrm{CFO}}[k]\mathbf{h}_{b,\mathcal{U}}[k]+\boldsymbol{\eta}[k],
\end{equation}
where $\boldsymbol{\Psi}_{\mathcal{U}}^{\mathrm{CFO}}[k] = \left[\boldsymbol{\Psi}_{\mathcal{U},1}[k]^T,\boldsymbol{\Psi}_{\mathcal{U},2}[k]^T\dots\boldsymbol{\Psi}_{\mathcal{U},M}[k]^T\right]^T \in \mathbb{C}^{MN_s \times G_R G_T}$ is the CFO-modified stacked sensing matrix.

Fig. \ref{new_simu}(a) compares the NMSE versus SNR performance for three CFO values, namely $\{-a\}$, $\{a\}$, and a random value picked from the interval $[-a,a]$, where $a=\frac{f_s}{2}$. From the figure, it is evident that the presence of CFO degrades the performance across the entire SNR range. This degradation is attributed to the progressive phase rotation introduced by CFO, which leads to a mismatch between the assumed signal model and the actual received signal. Consequently, the coherence of the measurements is disrupted, distorting the underlying beamspace sparsity structure and resulting in increased NMSE.

As seen in earlier discussions, the array response vectors are inherently frequency-dependent due to the beam-squint effect. Conventional models that employ frequency-independent steering vectors neglect this behavior, leading to a mismatch between the assumed and actual channel representations. Consequently, such models result in channel modeling mismatch which leads to performance degradation in practical implementations. Fig. \ref{new_simu}(b) illustrates the NMSE versus SNR performance for accurate and mismatched channel models. The performance degradation observed under mismatch is attributed to the discrepancy between the assumed and actual array response vectors, which results in model mismatch during channel estimation. Note that the performance comparison is carried out between the true and mismatched array response vectors, as given in Eq. \eqref{arrayresponsevector} and Eq. \eqref{arr_res_fc}, respectively.}
\vspace{-3mm}
\subsection{Analysis of Hybrid Transceiver Design}
Fig. \ref{new_simu}(c) illustrates the normalized array gain for the first and third users with respect to their physical steering directions. In the left plot, the uncompensated beam patterns show clear misalignment due to the beam-squint effect, where the beam directions vary across the frequency components, resulting in degraded gain at the intended angles. In contrast, the right plot demonstrates the performance improvement after applying the proposed TTD-based hybrid transceiver design. The frequency-dependent delays introduced by the TTD elements effectively compensate for the squint, aligning the beams more accurately with the user's physical directions, specifically at $\theta_1 \approx 102^\circ$ and $\theta_3 \approx 127^\circ$. Fig. \ref{SE}(a) depicts a substantial improvement in the achievable data rate when compared to existing state-of-the-art beamforming schemes which includes DPP \cite{dai2022delay}, SOMP \cite{el2014spatially}, interference cancellation (IC) \cite{li2016robust} across varying SNR levels. The performance of SOMP and IC degrades significantly in the presence of beam-squint effects, as neither of these approaches incorporates a delay-compensation mechanism. Without a TTD-based transceiver design, their beamformers become frequency-dependent, leading to beam misalignment and reduced array gain across the bandwidth. The DPP method also suffers from performance limitations due to the absence of low-resolution modeling via Bussgang decomposition, making it less effective under coarse quantization, which is common in practical THz systems. Furthermore, the DPP-SOMP attempts to enhance performance by estimating the angles using the SOMP algorithm and then compensating for beam-squint. However, this two-step approach introduces additional complexity and suffers from sensitivity to angular estimation errors due to its dependence on SOMP, which is inherently grid-based and prone to mismatch. In contrast, the proposed framework directly leverages the angular information embedded in the estimated beamspace, thereby eliminating the need for separate angle recovery and enabling an integrated TTD-based hybrid transceiver design.

{ Note that, as discussed in Section-\ref{ins_l_ss}, TTD elements effectively mitigate the beam-squint effect; however, they introduce hardware impairments in the form of insertion loss, which increases with delay and circuit complexity. This insertion loss leads to attenuation of the transmitted signal, thereby degrading the achievable SE. Fig. \ref{SE}(b), illustrates the impact of insertion loss on SE for both Rect-PSF and RRC pulse shaping. It is observed that SE decreases monotonically with increasing insertion loss, as the attenuation introduced by TTD elements limits the useful signal power available for reliable detection. Fig. \ref{SE}(c) illustrates the SE achieved for ideal and quantized delays. The SE achieved with quantized TTD delays is consistently lower than that with exact delays across the entire SNR range, reflecting the performance degradation due to delay quantization. This gap arises from the mismatch between the desired and realizable delays, which limits beamforming accuracy.}

Fig. \ref{SE}(d) illustrates the achievable data rate (in Gbps) between the proposed beamforming scheme and existing approaches, including DPP \cite{dai2022delay}, SOMP \cite{el2014spatially} under both RRC-PSF and Rect-PSF based dual-wideband channel models \cite{dovelos2021channel}. The Rect-PSF channel tends to introduce spectral leakage, resulting in inter-symbol interference and reduced spectral confinement. In contrast, while the RRC-PSF channel provides better spectral shaping, it suffers from edge attenuation effects, which can degrade signal recovery at the receiver. This highlights a fundamental \textit{trade-off} between both the dual-wideband channel formulations. Moreover, the proposed method consistently outperforms all baseline techniques, demonstrating its robustness and its ability to effectively mitigate the beam-squint effect. Note that, all the results are obtained for $4$-bit ADC resolution. Thus, the results clearly demonstrate the effectiveness of the proposed framework in jointly addressing channel estimation and beamforming challenges, offering robust mitigation of wideband distortions such as beam-squint effect, while remaining efficient under practical low-resolution hardware constraints. { Potential future work includes extending the proposed framework to high-mobility scenarios to investigate its performance under time-varying channel conditions.}
\vspace{-3mm}
\section{Conclusion} \label{conclusion}
\vspace{-1mm}
Variational Bayesian inference-based channel estimation framework for wideband THz MIMO systems is proposed in the work. To accurately capture the wideband propagation characteristics, we formulated a dual-wideband channel model incorporating RRC-PSF. Bussgang decomposition was employed to linearize the nonlinear signal model arising from low-resolution ADCs, enabling tractable inference under coarse quantization. The proposed channel estimation scheme was developed and analyzed under both on-grid and off-grid angular assumptions, offering robustness against basis mismatch and improved angular resolution. Leveraging the estimated beamspace information, we then designed a hybrid transceiver based on TTD to mitigate the beam-squint effect and enable frequency-flat beam alignment across subcarriers. Simulation results demonstrate that the proposed approach consistently outperforms existing techniques in both channel estimation accuracy and achievable data rate, while remaining efficient under low-resolution hardware constraints.
\appendices
\vspace{-3mm}
{ \section{Calculation of Noise Covariance} \label{appendix-a}}
\vspace{-1mm}
The received signal after performing analog combining and before passing through the low-resolution ADC can be given as
\vspace{-2mm}
\begin{align}
    \tilde{\mathbf{y}}_m(p) = \mathbf{W}_{\mathrm{RF},m}^H \sum_{u=1}^U \mathbf{H}_{p,u} \otimes_K & \big(\mathbf{F}_{\mathrm{RF},m,u} \mathbf{F}_{\mathrm{BB},m,u} \boldsymbol{\mathfrak{P}}_{m,u}^{(p)}\big) + \notag \\ & \mathbf{W}_{\mathrm{RF},m}^H \tilde{\mathbf{v}}_m(p). \label{before-ADC}
\end{align}
Let $\mathbf{x}_{m,u}(p) = \mathbf{F}_{\mathrm{RF},m,u} \mathbf{F}_{\mathrm{BB},m,u} \boldsymbol{\mathfrak{P}}_{m,u}^{(p)}$ for notational simplicity. Therefore, $\mathbb{E}(\mathbf{x}_{m,u}(p)\mathbf{x}_{m,u}^H(p)) = \sigma_b^2 \mathbf{F}_{\mathrm{RF},m,u} \mathbf{F}_{\mathrm{BB},m,u} \mathbf{F}_{\mathrm{BB},m,u}^H \mathbf{F}_{\mathrm{RF},m,u}^H = \mathbf{R}_{xx,m}$. Expanding the circular convolution as defined in Section-I(C) of the main paper, Eq. \eqref{before-ADC} can be re-written as
\vspace{-2mm}
\begin{align}
    &\tilde{\mathbf{y}}_m(p) = \mathbf{W}_{\mathrm{RF},m}^H\sum_{u=1}^U[\mathbf{H}_u(0)\mathbf{x}_{m,u}(p)+ \mathbf{H}_u(1)\mathbf{x}_{m,u}(p-1)+ \notag \\ & \cdots+ \mathbf{H}_u(K-1)\mathbf{x}_{m,u}(p-K+1)] + \mathbf{W}_{\mathrm{RF},m}^H\tilde{\mathbf{v}}_m(p). \label{expand_ADC}
\end{align}
Let $\tilde{\mathbf{C}}_m \in \mathbb{C}^{N_{\mathrm{RF}}^R \times N_{\mathrm{RF}}^R}$ represents the covariance matrix which is defined as $\tilde{\mathbf{C}}_m = \mathbb{E}\{\tilde{\mathbf{y}}_m(p)\tilde{\mathbf{y}}_m^H(p)\}$. After substituting $\tilde{\mathbf{y}}_m(p)$ into the expression of the covariance matrix, we obtain
\vspace{-2mm}
\begin{align}
    &\tilde{\mathbf{C}}_m = \mathbf{W}_{\mathrm{RF},m}^H \sum_{u=1}^U[\mathbf{H}_u(0)\mathbf{R}_{xx,m}\mathbf{H}_u^H(0)+\cdots+ \notag \\ & \mathbf{H}_u(K-1)\mathbf{R}_{xx,m}\mathbf{H}_u^H(K-1)]\mathbf{W}_{\mathrm{RF},m}^H + \sigma_n^2\mathbf{W}_{\mathrm{RF},m}^H\mathbf{W}_{\mathrm{RF},m}.
\end{align}
Therefore, the received signal covariance matrix before passing through low-resolution ADCs can be expressed as $\tilde{\mathbf{C}}_m = \mathbf{W}_{\mathrm{RF},m}^H\mathbf{Q}_m\mathbf{W}_{\mathrm{RF},m} + \sigma_n^2\mathbf{W}_{\mathrm{RF},m}^H\mathbf{W}_{\mathrm{RF},m}$ where $\mathbf{Q}_m = \sum_{u=1}^U \sum_{n=0}^{K-1}\mathbf{H}_u(n)\mathbf{R}_{xx,m}\mathbf{H}_u^H(n)$. Additionally, the noise covariance matrix after passing through low-resolution ADC is given as $\mathbf{C}_m = \kappa(1-\kappa)\mathrm{diag}(\tilde{\mathbf{C}}_m)$.
\vspace{-2mm}
{ \section{Condition for maximum array gain} \label{appendix-b}}
\sloppy
Revisiting the array response vector Eq. \eqref{arrayresponsevector} and normalized array gain Eq. \eqref{normalizedgain} and substituting Eq. \eqref{arrayresponsevector} in \eqref{normalizedgain}, one obtains $\tilde{\Omega} = \frac{1}{N_{R}^{l}} \tilde{\mathrm{ar}}^H \tilde{\mathrm{af}}$ where $\tilde{\mathrm{ar}} = \frac{1}{N_{R}^{l}}[1, e^{-j \pi \vartheta_k \sin \theta_k}, \cdots, e^{-j \pi (N_{R}^l-1) \vartheta_k \sin \theta_k}]^T$ and $\tilde{\mathrm{af}} = [1, e^{-j \pi \sin \theta}, \cdots, e^{-j \pi \left(N_{R}^l-1 \right) \sin \theta}]^T$. Further simplifying the above equation and assuming $\theta$ is small, we obtain
\vspace{-2mm}
\begin{align}
    \tilde{\Omega} = \frac{1}{N_{R}^l}\Big|\sum_{n=0}^{N_R^l-1}e^{jn\pi(\vartheta_k \theta_k - \theta)} \Big|.
\end{align}
Therefore, in order to achieve maximum array gain, $\theta_k = \frac{\theta}{\vartheta_k}$ needs to be true. 
\vspace{-4mm}
{ \section{Calculation of optimal angle} \label{appendix-c}}
Revisiting Eq. \eqref{normalizedgain} and substituting the array response vector $\tilde{\mathbf{a}}_k = \mathrm{blkdiag}\big(\tilde{\mathbf{a}}_1^T, \cdots, \tilde{\mathbf{a}}_S^T\big)e^{-j2 \pi f_k\mathbf{t}}$ in the normalized array gain expression, one obtains
\vspace{-2mm}
\begin{align}
    \tilde{\Omega}_k = \big|(\tilde{\mathbf{a}}(\theta,f_c))^H \tilde{\mathbf{a}}(\theta_k,f_k)\big|,
    \end{align}
\begin{figure*}[htbp]
    \begin{align}
    &\tilde{\Omega}_k  = \Big|\frac{1}{N_R^l}  \Big\{\big(1+  e^{j \pi \theta}e^{-j\pi\vartheta_k\theta_k}+  \cdots+e^{j \pi (Z-1)\theta}e^{-j\pi(Z-1)\vartheta_k\theta_k}\big) +  e^{-j\pi\upsilon_k} \big(1+ \cdots +e^{j \pi (2Z-1)\theta}e^{-j\pi(2Z-1)\vartheta_k\theta_k} \big)  +  \notag \\ & \cdots + e^{-j\pi (S-1) \upsilon_k} \big(1+e^{j \pi (S-1)Z \theta}e^{-j\pi (S-1)Z \vartheta_k\theta_k}+ \cdots +  e^{j \pi (SZ-1)\theta}e^{-j\pi(SZ-1)\vartheta_k\theta_k} \big) \Big\} \Big|, \label{normalize}
\end{align}
\vspace{-5mm}
\begin{align}
    \tilde{\Omega}_k  = \frac{1}{N_R^l} \Big| \sum_{s=1}^S \sum_{z=1}^Z e^{j \pi [(s-1)Z+(z-1)]\theta} e^{-j \pi \vartheta_k[(s-1)Z+(z-1)]\theta_k} e^{-j\pi(s-1)\upsilon_k}\Big|, \label{normalize_angle}
\end{align}
\hrulefill
\vspace{-1.6\baselineskip}
\end{figure*}
which can be further simplified to Eq. \eqref{normalize}-\eqref{normalize_angle} and finally yields
\vspace{-2mm}
\begin{align}
    \tilde{\Omega}_k  = \frac{1}{N_R^l} \Big| \sum_{s=1}^S e^{j\pi(s-1)\left\{Z\theta - \vartheta_k Z \theta_k -\upsilon_k \right\}}\sum_{z=1}^Z e^{-j\pi(z-1)\left\{-\theta+\vartheta_k\theta_k \right\}} \Big|. \notag
\end{align}
Therefore, the normalized array gain can be represented in the form of Dirichlet sinc function, as
\vspace{-2mm}
\begin{align}
    \tilde{\Omega}_k = \frac{1}{N_R^l} \big|\Xi_S\big(Z\theta - \vartheta_k Z \theta_k -\upsilon_k \big) \Xi_Z \big(-\theta+\vartheta_k\theta_k \big)\big|.
\end{align}
Thus, the optimal angle at which the array gain will be maximum can be obtained by substituting $Z\theta - \vartheta_k Z \theta_k -\upsilon_k = 0$ which finally yields Eq. \eqref{optimal_angle}.
\vspace{-1 \baselineskip}
\bibliographystyle{IEEEtran}
\bibliography{References}
\end{document}